\documentclass[footinbib,a4paper,aps,prb,amsmath,amssymb,showpacs,10pt,twocolumn,floatfix]{revtex4}
\usepackage{graphicx}
\usepackage{graphics,color}
\begin{document}
%\draft
%
\title{Universality class of replica symmetry breaking, scaling behavior, and the low-temperature fixed-point order function of the Sherrington-Kirkpatrick model}
\author{R. Oppermann, M.J. Schmidt}
\address{Institut f. Theoretische Physik, Universit\"at
W\"urzburg, Am Hubland, 97074 W\"urzburg, FRG}
\date{\today}
\begin{abstract}
A scaling theory of replica symmetry breaking (RSB) in the SK-model is presented in the framework of critical phenomena for the scaling regime of small inverse RSB-orders $1/\kappa$, small temperatures $T$, and small magnetic fields $H$. We employ the pseudo-dynamical picture (\prl98, 127201 (2007)) with two critical points ${\cal CP}1$ and ${\cal CP}2$ where separated temperature- and magnetic field-scaling is obtained near the order function's pseudo-dynamical limits $lim_{_{a\rightarrow\infty}}q(a)=1$ and $lim_{_{a\rightarrow0}}q(a)=0$ at $(T=0,H=0)$. An unconventional scaling hypothesis for the free energy is given, modeling this separated scaling in accordance with detailed numerical self-consistent solutions for up to $200$ orders of RSB.
Divergent correlation-lengths $\xi_{{\cal CP}1}(T)\sim T^{-\nu_{_{\hspace{-.03cm}T}}}$ and $\xi_{{\cal CP}2}(H)\sim H^{-\nu_{_{\hspace{-.03cm}H}}}$ describe the RSB-criticality as a long-range correlation effect occurring on the pseudo-lattice of RSB-orders. Rational-valued exponents $\nu_{_T} =3/5$ and $\nu_{_H}=2/3$ are concluded with high precision from high-order RSB scaling (in analogy with finite size scaling) and using a new fixed point extrapolation method.
Power laws, scaling relations, and scaling functions are analyzed.
Near ${\cal CP}1$, the non-equilibrium susceptibility is found to decay like $\chi_1=\kappa^{-5/3}f_{1}(T/\kappa^{-5/3})$, the $T=0$-entropy like $S\sim\chi_1^2$,
while ($T$-normalized) Parisi box sizes diverge like $a_i=\kappa^{5/3} f_{a_i}(T/\kappa^{-5/3})$,
with $f_{1}(\zeta)\sim\zeta$ and $f_{a_i}(\zeta)\sim1/\zeta$ for $\zeta\rightarrow\infty$, $f(0)$ finite.
Near ${\cal CP}2$, where the magnetic field $H$ controls the critical behavior
(while temperature is irrelevant), a power law $H^{2/3}$ is retrieved for plateau-height
%(and -width)
of the order function $q(a)$ according to $q_{pl}(H)=\kappa^{-1}f_{pl}(H^{2/3}/\kappa^{-1})$ with $f_{pl}(\zeta)|_{\zeta\rightarrow\infty}\sim\zeta$ and $f_{pl}(0)$ finite.
The order function $q(a)$ links ${\cal CP}1$ with ${\cal CP}2$ and is obtained as a fixed point function $q^*(a^*)$ of RSB-flow, in agreement with integrated fixed-point energy and susceptibility distributions.
Similarities with directed polymers in $1+1$ dimensions, with $d=1$ solution and Flory-Imry-Ma type solutions of the KPZ-equation are discussed.
\end{abstract}
\maketitle
%XXXXXXXXXXXXXXXXXXXXXXXXXXXXXXXXXXXXXXX introduction XXXXXXXXXXXXXXXXXXXXXXXXXXXXXXXXXXXXXXXXXXX
\section{Introduction}
%XXXXXXXXXXXXXXXXXXXXXXXXXXXXXXXXXXXXXXXXXXXXXXXXXXXXXXXXXXXXXXXXXXXXXXXXXXXXXXXXXXXXXXXXXXXXXXXX
%
The far-reaching usefulness of spin glass theories \cite{MPV,apyoung-book,parisi-book} and of its key structural elements such as frustration, disorder, hierarchical order, ultrametricity, complexity, freezing transition, is witnessed well by applications entering even life-sciences and trans-disciplinary research fields. Physical models, where these key structures acquired a specific mathematical meaning, find very broad applications beyond their origin in frustrated magnetism. Let us mention, apart from fields like neural networks, computer science, and econophysics, the fascinating sociological applications to opinion- and group dynamics \cite{s.galam1,s.galam2}, biological applications to RNA-folding \cite{f.david,p.higgs,laessig-wiese} including the quantum chromodynamical analogy and random matrix theory\cite{zee-orland}.
It seems natural to search for universal features of unifying models both in the general sense and in the precise meaning of the renormalization group\cite{f.david}.

The 3SAT optimization problem and its close relation with the $T=0$ Sherrington-Kirkpatrick model \cite{SK} or RNA-folding in biophysics \cite{p.higgs}, where glass transitions exist within the secondary RNA-structure \cite{emarinari,mmueller,f.david}, provide examples where even the zero temperature limit is either exact or close to the realistic situation.
In physics, spin glass phases are usually confined to a low temperature regime and some applications are rather remote from it. Yet, knowing the ground state structure remains important.
For one of the most fertile standard models, the Sherrington-Kirkpatrick model \cite{SK} (SK-model), the hierarchical ground state structure is meanwhile confirmed \cite{Talagrand} as predicted by Parisi a long time ago \cite{Parisi3,Parisi1}. Explicit analytic solutions on the other hand or meaningful approximations are still required. They may lead to improved understanding and could be potentially fruitful for progress in more complicated (non-mean-field finite-range, quantum-) models.

The attempt to link the SK-model behavior deep inside its ordered (spin glass) phase with the theory of critical phenomena may appear unmotivated at first sight, since the infinite-ranged spin interaction suggests 'only' mean-field behavior.
However the SK-model is not simple below its mean-field transition. Its replica theory \cite{binder-young} allows to imagine how the Ising Hamiltonian with infinite-ranged random interaction can become potentially critical, when it is dressed-up with the hierarchical order parameter structure in the replica symmetry broken (RSB) phase\cite{parisi-book}.
We shall argue in this paper that, as the number of tree levels of this hierarchical structure grow to infinity, a particular correlation length between the tree levels can be defined which diverges as $T\rightarrow0$. It allows to describe critical behavior due to the accumulated effect of ever finer structures at the highest tree level. This property specifies a kind of universality class, which helps to compare with similar behavior in different physical systems and in other scientific areas like biology, sociology, (mathematical) psychology as well, where evidently frustrated random (and in cases range-free) interactions are important.

Nonanalytic power laws (with rational exponents) for the SK-model had been discussed in many different respects, as for example the finite-size cutoff (or finite spin number) dependence \cite{boettcher,bouchaud-energy-exponents,mikemoore}. One may also mention the exponent of the Almeida Thouless line\cite{binder-young}. However a link to specific critical points was not made.
%and the question whether some power laws simply hold everywhere in the ordered phase remained open.

In the present paper, we shall report progress in understanding replica symmetry breaking in the Sherrington-Kirkpatrick model \cite{SK} as a critical phenomenon; this refers to scaling behavior on one hand and to (numerically determined) fixed point functions under RSB-order flow $\kappa\rightarrow\infty$ on the other. Nonanalytic scaling behavior is described as a function of the inverse RSB-order decreasing to zero either together with temperature $T\rightarrow0$, or together with the external magnetic field. Temperature- and field-scaling are well separated and reside in opposite limits of a pseudo-dynamic variable $1/a$ (see Ref.\onlinecite{prl2007}), as sketched by Fig.\ref{fig:scaling-variables}
\footnote{in addition to $1/\kappa$ we also consider scaling with respect to the pseudo-dynamic variable $1/a$, treating both as quasi-continuous scaling variables - one may imagine the analogy of a large enough lattice such that the discreteness of momenta can be neglected.}.

Critical phenomena are in general categorized by universality classes and described by criteria like global symmetries. Certain details (on shorter range) become irrelevant and suppressed in the regime of divergent correlation lengths.
In the early years of the development of phase transition theory and critical phenomena, Kadanoff's initializing ideas of universality and rescaling, Stanley's scaling theory, and Wilson's renormalization group led to the modern understanding of critical behavior \cite{RG-review}.
In recent years the functional renormalization group was advocated to understand better disorder-related criticality \cite{wiese}.

Freezing transitions into spin glass phases were analyzed in renormalization framework too, the ordered phase itself remained however mysterious, in particular for the non-mean-field models.
In a famous work on scaling in spin glasses D. Fisher and Sompolinsky \cite{dfisher-hsompolinsky} explained the complications of mean field models (or mean field regimes of finite-range spin glasses above $d=6$ and $d=8$) and the multiple violations of scaling relations. In particular they mentioned the violation of temperature- versus magnetic field scaling within the ordered phase. In a different manner, we re-encounter this problem and explain a certain decoupling of field- from temperature-scaling by the presence of two different critical points of RSB in the low temperature limit.

Crucial questions like the relationship between Parisi's RSB and the Fisher-Huse droplet theory \cite{fisher-huse} of the ordered phase of real spin glasses (or their reconciliation) became - since a long time but perhaps currently with more good reason - a field of intensive research\cite{cyrano,monthus-garel-condmat07-dec}.
Since droplet theory is interpreted to govern the ordered phase by a $T=0$ fixed point, it appears very desirable to understand RSB as a $T=0$ fixed point theory too. Attempts have been independently undertaken by several authors and also in different fields of application, as the examples in Refs.\onlinecite{wiese,prl2005,pankov-prl,mm-pankov,prl2007} show.

The latter point is elaborated in the present article. Despite the mean field character of the SK-model, RSB introduces apparently nonanalytic critical behavior of one-dimensional type (the unbroken replica-symmetric solution does not show any of these phenomena) together with special diverging correlation lengths. The challenge to handle RSB-effects correctly and to make the SK-solution a fruitful basis for real physical applications led us to a scaling theory intimately based on extreme high order numerical results.

In previous publications \cite{prl2007,prl2005} we reported the existence of two critical points and of discrete spectra which survived in the limit of infinite replica symmetry breaking ($\infty$-RSB) for the SK-model at $T=0$, perhaps surprising since the $\infty$-RSB limit is in generally known only as the 'continuum limit'. Indeed, a continuum scaling theory, dealing with the $T\rightarrow 0-$limit at $\kappa=\infty$) was published by Pankov \cite{pankov-prl} recently. Its role and limitation to the temperature-controlled critical point ${\cal CP}1$ has been addressed in our previous publication \cite{expcpaper} together with a comparison of our work with the much older so-called PaT-scaling \cite{PaT}. In the present article we do neither use Pankov- nor PaT-scaling, but construct a different scaling approach, which includes RSB-order-scaling, and is exclusively guided by the theory of critical phenomena. In accordance with previous (naive) functional renormalization group arguments \cite{prl2005} we analyze the approach to full RSB formation ($\kappa\rightarrow\infty$) not only at $T=0$ but also in the $(H,T)$-plane for small values of temperature $T$ and magnetic field $H$ and, of course, as a function of RSB-order.
(neither real space nor real-time space are involved as a consequence of the SK-model's nature).

We suppose here that RSB orders, counted by integers $1,2,...,\kappa$, can be viewed as equidistant sites forming a pseudo-lattice. In analogy with a real-space lattice, which needs to be infinitely large in order to allow diverging correlation lengths and hence support critical phenomena, the pseudo-length cutoff $\kappa$ must be sent to infinity.
The known fact that increasingly high orders of RSB are needed (for good approximations) as the temperature decreases towards zero implies the role of $T$ as an effective cutoff of nonanalytic behavior in the RSB-limit ($T$ playing the role of a symmetry breaking relevant perturbation in standard critical phenomena).
Thus it also inherits the idea of scaling RSB-order $\kappa$ with temperature $T$. Conversely, a maximum RSB order $\kappa$ serves as a cutoff of criticality. A speciality of RSB is that it appears in the shape of a pseudo-dynamical critical phenomenon \cite{prl2005,prl2007}, which recalls the celebrated dynamical representation of Sompolinsky \cite{sompolinsky}. A technically important difference however being the absence of a stochastic field, which we reserve for more complicated couplings to faster degrees of freedom \cite{pssc2007}.

A scaling theory, near $T=0$ in particular, is important for several different reasons. First, it expresses the numerically determined features of the SK-model in a universal form, which helps to identify model-independent features and places the SK-model and its RSB in a wider context. Let us mention that directed polymers (or for example the queuing transition and the totally asymmetric exclusion process \cite{k.johansson,queuing-denNijs}, or certain partial differential equations, involve rational exponents as multiples of $1/3$ too).
The scaling theory also puts constraints on the shape of an effective field theory. It has the virtue of isolating critical features which must be represented correctly by an effective theory that simplifies the SK-model.
\footnote{The crossover behavior (between critical points) may not need an exact representation.}
The simpler theory should allow to control generalizations to finite range or other complications.
The scaling theory offers also a special look on eventual scenarios of an RSB breakdown, as it may occur due to finite range interactions. The collapse of finite $T_c$ below a lower critical dimension, will eventually combine RSB-criticality with the freezing transition as $T_c=0$.\\
%The mapping to a field theory which becomes as simple as a Langevin differential equation in the SK %model limit \cite{pssc2007}, has still to be completed.\\

%XXXXXXXXXXXXXXXXXXXXXXXXXXXXXXXXXXXXXXXXXXXXXXXXXXXXXXXXXXXXXXXXXXXXXXXXXXXXXXXXXX
{\it The paper is organized as follows:}\\
%XXXXXXXXXXXXXXXXXXXXXXXXXXXXXXXXXXXXXXXXXXXXXXXXXXXXXXXXXXXXXXXXXXXXXXXXXXXXXXXXXX

Section \ref{scaling-concept} and \ref{RSB-correlation-length} describe the basic elements of the present scaling theory. The spaces spanned by the scaling variables at both critical points are described in section \ref{scaling-concept}. In section \ref{RSB-correlation-length} a correlation length is introduced on the pseudo-lattice of RSB-orders (to our best knowledge for the first time) and, anticipating the self-consistent numerical results of the following Sections \ref{fixed-point-function}-\ref{U-distribution} (for details see Ref.\onlinecite{expcpaper}), the role of finite temperatures (or finite magnetic fields) as soft cut-offs of the divergence of this correlation length is explained.

Section \ref{fixed-point-function} demonstrates how the order parameter function $q(a)$ can be regarded and obtained as a fixed point function $q^*(a^*)$ under RSB-flow $\kappa\rightarrow\infty$.

Section \ref{finite-T-scaling} includes and combines finite temperature scaling near critical point ${\cal CP}1$ with RSB-order scaling. Scaling functions are obtained, which fit the detailed data of $200$ RSB-orders, and explain the non-commuting singular limit $\kappa\rightarrow\infty, T\rightarrow 0$. In a similar way, \ref{finite-H-scaling} includes magnetic field scaling near ${\cal CP}2$.

In Section \ref{free-energy-scaling} we present unconventional scaling-contributions to the free energy, to the entropy, and internal energy, which are compatible with the numerical self-consistent solutions.
%Unlike length-scaling in real space there is no extensivity constraint of $F$ on the $\kappa$-dependence, with %the RSB-order $\kappa$ playing the role of pseudo-length.

In Section \ref{U-distribution} the ground state energy distribution is given as a function of pseudo-time and also shown as a function of the (normalized) Parisi levels $l/\kappa$ such that the flow towards a energy-per-level fix point function results as the RSB-order tends to infinity.

In Section \ref{pseudo-dynamical-scaling} we finally consider pseudo-dynamic scaling of the order function $q(a)$ in the vicinity of both critical points before concluding with details of $q(a)$ as revealed by its derivatives in Section \ref{q(a)-derivatives}.

%
%XXXXXXXXXXXXXXXXXXXXXXXXXXXXXX             1              XXXXXXXXXXXXXXXXXXXXXXXXXXXXXXXXX
\section{The scaling scenario}
\label{scaling-concept}
\begin{figure}[here]
\resizebox{.5\textwidth}{!}{%
\includegraphics{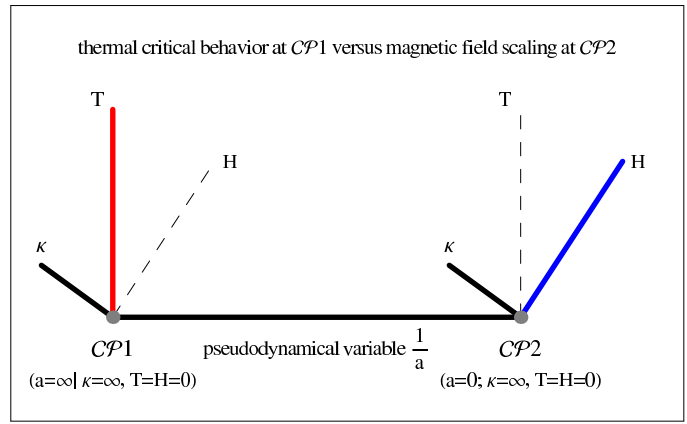}}
\caption{\label{fig:scaling-variables}Scaling regimes near the two critical points of the SK-model at zero temperature, zero field, and infinite replica symmetry breaking. The critical points are separated by the full range of continuously distributed pseudo-times $1/a$.}
\end{figure}

%XXXXXXXXXXXXXXXXXXXXXXXXXXXXXXXXXXXXXXXXXXXXXXXXXXXXXXXXXXXXXXXXXXXXXXXXXXXXXXXXXXXXXXXXXXX
We introduce the (RSB-)scaling idea by viewing the formation of full RSB as a critical phenomenon with two critical points in the pseudo-dynamic limits $a=0$ and $a=\infty$ at $T=0$, $H=0$. We do not a priori impose a relationship between the two critical points, but consider the pseudo-dynamical crossover between them by means of the order function $q(a)$ on $0\leq a\leq\infty$. Fig.\ref{fig:scaling-variables} illustrates the relative position of the two critical points and the sets of scaling variables near these points.

In particular one may notice that the dynamical variable $1/a$ and the RSB-order $\kappa$ define a $(1+1)$-dimensional analogy of problems with one time- and one real space dimension.
Since the free energy or internal energy are integrals over all pseudo-times, as for example given below in Eqs.(\ref{eq:U}),(\ref{eq:F}), we do not start from a single scaling hypothesis for the free energy $F$.
Instead we construct the scaling hypothesis for each of the two different scaling contributions, originating in these separated critical points and. As Fig.\ref{fig:scaling-variables} shows, a different set of scaling variables should be used in order to match the numerical results.

It is remarkable that temperature- and magnetic field-scaling become decoupled, because they belong to different scaling regimes.
Scaling with the respect to the order $\kappa$ of RSB measures the approach of the equilibrium solution at $\kappa=\infty$ (full RSB) and therefore can be viewed as a kind of non-equilibrium dynamics (in the sense that each finite order is unstable towards higher RSB-orders).
Thus an element of dynamic scaling is contained. Using the pseudo-time $1/a$ as an additional scaling variable, we analyze the order function $q(a; T,H)$ and its pseudo-dynamic scaling behavior. A dynamic crossover between the two critical points ${\cal CP}1$ and ${\cal CP}2$ is then described by means of $q(a)$. Moreover, the order function is evaluated as a fixed point function of the RSB-flow letting $\kappa\rightarrow\infty$.

The present scaling theory is then fitted to high precision numerical data, which were obtained recently for the Sherrington-Kirkpatrick model given by the Hamiltonian
$${\cal H}=\sum_{i<j} J_{ij}s_i s_j-H\sum_i s_i$$
with quenched, infinite-ranged, and Gaussian-distributed random couplings $J_{ij}$ (with variance $J^2/N$) between classical spins $s_i=\pm1$.
The method was described in Ref.\onlinecite{expcpaper} and will not be described again in this article. It allowed not only to go beyond earlier high-order studies \cite{prl2007}, but contained also new analytical elements.
As a consequence we are able to predict the values of critical exponents, evaluate amplitudes, calculate analytical models of various scaling functions including cases with very singular crossover.

The numerical material includes the self-consistent solutions in all orders of RSB up to\\
i) the current maximum of $\kappa=200$ RSB at $T=0$ and $H=0$,\\
ii) $50$ orders for a dense grid of finite temperatures in the range $0\leq T\leq 0.3$ for $H=0$, and\\
iii) $20$ orders of RSB for a dense grid of finite magnetic fields $0\leq H\leq 0.5$ at zero temperature.\\

{\it We note that all energies are given in units of $J$}.

%XXXXXXXXXXXXXXXXXXXXXXXXXXXXXXXXXXXXXXXXXXXXXXXXXXXXXXXXXXXXXXXXXXXXXXXXXXXXXX
\section{Divergent correlation length on the one-dimensional pseudo-lattice of RSB-orders}
\label{RSB-correlation-length}
%XXXXXXXXXXXXXXXXXXXXXXXXXXXXXXXXXXXXXXXXXXXXXXXXXXXXXXXXXXXXXXXXXXXXXXXXXXXXXX
The high order self-consistent solutions led us to consider a pseudo-lattice of RSB-orders with unit lattice constant. The maximal order $\kappa$, for which self-consistent results have been obtained numerically, can be viewed as a (sharp) pseudo-length cutoff. In analogy with the well-known finite size scaling of critical phenomena, one may consider scaling by varying this finite maximum RSB-order $\kappa$. Naturally this defines a one-dimensional problem without translational invariance though. Moreover it is known that finite temperatures or finite fields serve each as a soft-cutoff for the maximal order of RSB, which is needed to obtain good approximations: the higher the temperature the less RSB orders are needed to obtain a certain quality of results\footnote{For example in order to guarantee a nonnegative entropy when temperature decreases towards zero, one must scale $\kappa$ up like $T^{\nu_T}$ such that the RSB-order stays larger than $\xi_{\kappa}(T)$}.

In other words, higher orders become uncorrelated because they only have weak and/or negligible effects. In this sense the correlation length of different RSB orders becomes cutoff by finite $T$ or $H$.
Anticipating our results below, this definition of the correlation lengths $\xi_{\kappa}(a,T,H)$ shows power law divergences with rational-valued exponents given by
\begin{equation}
\label{nu-T}
\xi_{{\cal CP}1}\equiv \xi_{\kappa}(a=\infty,H=0,T)\sim T^{-\nu_{T}},\quad \nu_T=3/5
\end{equation}
\begin{equation}
\label{nu-H}
\xi_{{\cal CP}2}\equiv \xi_{\kappa}(a=0,H,T=0)\sim H^{-\nu_H}, \quad \nu_H=2/3
\end{equation}

In the chapters below, we shall find scaling functions of the form $f(\kappa/\xi_{\kappa})$.
Apparently the correlation length exponent $\nu_H$ cannot be simply related to $\nu_T$, since in the vicinity of one of the two critical points one finds either $T$ and no $H$-dependence (${\cal CP}1$) or vice versa (${\cal CP}2$) (unlike conventional scaling where $T\sim H^{1/\beta\delta}$).
The power laws (1) and (2) should also hold for $a\gg\xi_0$ and $a\ll\xi_0$ respectively, where $\xi_0$ was introduced in Ref.\onlinecite{prl2005} as a finite characteristic length ($\approx 1.13$) which sets the scale for crossover from almost linear regime, $q\sim a$, through a maximal-curvature crossover ($a\approx \xi_0$) to $1-q\approx 1/a^2$ behavior of the order function $q(a,T=0,H=0)$.

Let us note some similarities with critical behavior of other systems. Very remarkable appears the fact that the exponent $\nu_T=\frac35$, describing the RSB-correlation-length divergence at ${\cal CP}1$, coincides with the value given by Garel and Orland for the variational {\it domain-wall solution} of $(1+1)-$dimensional directed polymers \cite{garel-orland}.

For general $d<2$ these authors reported $\nu=3/(d+4)$, while a second solution called domain-solution yields $\nu=1/(4-d)$. Hence, in $(1+1)$-dimensional case of directed polymers, these two solutions give $\nu=3/5$ and $\nu=1/3$ respectively.  If $H^2$ in our case could be scaled like $T$ then our second correlation length exponent $\nu_H$ would agree with the domain solution for directed polymers (DP). Moreover, the roughness-exponent of the DP in $1+1$ dimension assumes precisely this value $2/3$. Of course, these are only hints, and universality classes can coincide accidentally at integer dimensions.
It appears however very tempting to analyze whether our two critical points ${\cal CP}1$ and ${\cal CP}2$, distinguished by the opposite pseudo-dynamic limits $a=\infty$ and $a=0$ respectively, find their counterparts in those domain- and domain-wall solutions of the directed polymers. Whether such a formal analogy holds (despite the infinite-range spin glass interaction), should probably be decided by comparison or mapping of the corresponding field theories. Searching for eventual relations in higher dimensions for finite range spin glass interactions is also an exciting question.
We shall come back to related questions and pseudo-dynamic scaling in section \ref{pseudo-dynamical-scaling}.\\

\begin{figure*}
\resizebox{.8\textwidth}{!}{%
\hspace{-1.2cm}
\includegraphics{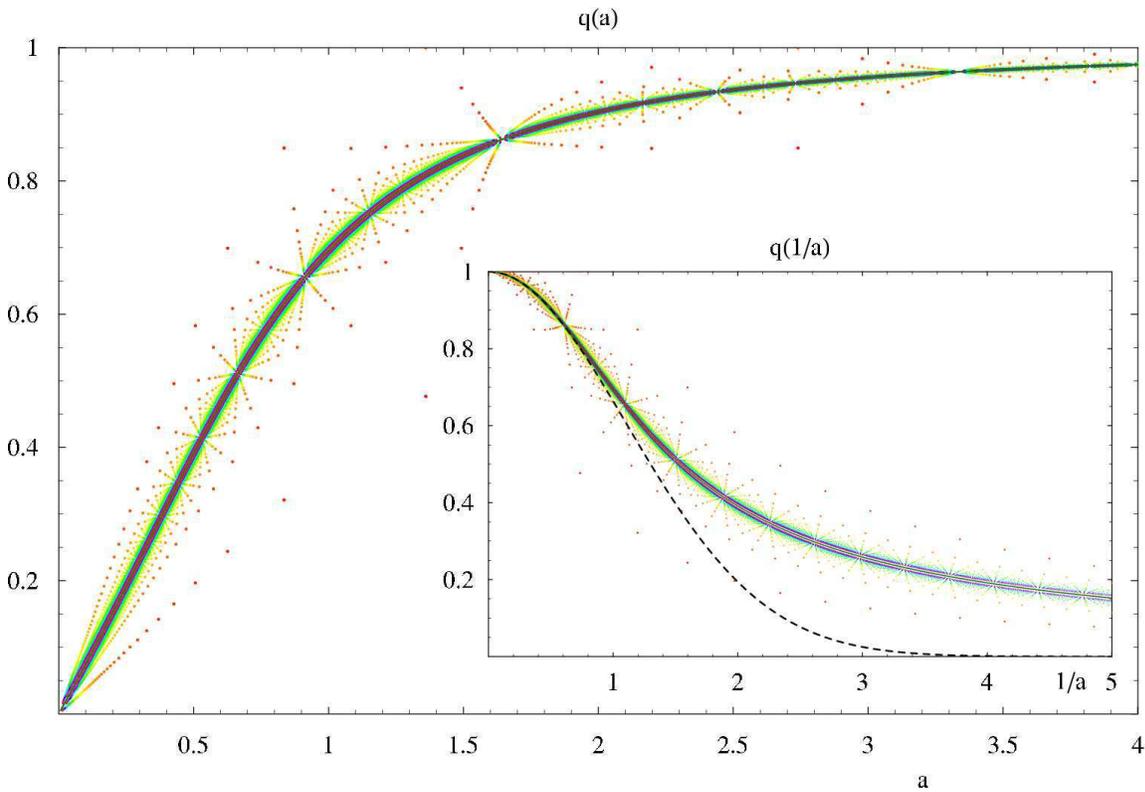}
}
\caption{\label{fig:q(a)-bounds}Main Figure and Insert present an (overlapping) wide-scale overview of all $(T=0,H=0)$ selfconsistent solutions $\{a_l^{(\kappa)}, q_{l+1}^{(\kappa)}\}$ and $\{a_l^{(\kappa)}, q_{l}^{(\kappa)}\}$, where $l=1...\kappa$ for each RSB-order $\kappa=1,2,...,200$. One color cycle from $\kappa=1$ (red, remote points) to the highly dense regime $\kappa=200$ (red, almost continuous curve) demonstrates how the true order function $q(a)$ is approached by the almost continuous $q_{l+1}(a_l)$ from below and by $q_l(a_l)$ from above.
The Insert shows the corresponding $\{1/a,q\}$-data and the function $q(1/a)$ they converge too.
The shape near $\tau\equiv \frac{1}{a}=0, ({\cal CP}1)$ is almost Gaussian (dashed black curve: $\exp(-0.41/a^2)$) but crosses over for moderate and small $1/a$ to strongly non-Gaussian and a tail behavior $q(\tau)\sim 1/\tau$.}
\end{figure*}
%
%
%XXXXXXXXXXXXXXXXXXXXXXXXXXXXXXXXXXXXXXXXXXXXXXXXXXXXXXXXXXXXXXXXXXXXXXXXXXXXXXXXXXXXXXXXXXXXX
\section{The $T=0$ order function as a fixed point function $q^*(a^*)$ in the RSB limit ($\kappa=\infty$)}
\label{fixed-point-function}
%XXXXXXXXXXXXXXXXXXXXXXXXXXXXXXXXXXXXXXXXXXXXXXXXXXXXXXXXXXXXXXXXXXXXXXXXXXXXXXXXXXXXXXXXXXXXXX
%
The idea of finding the $T=0$ order function as a fixed point function in the RSB-limit arose from renormalization group arguments as designed in Ref. \onlinecite{prl2005}. It reemerges now in a literally obvious way when we plot in Figure \ref{fig:q(a)-bounds} the whole set of numerical self-consistent solutions
$\{a_l^{(\kappa)}, q_{l+1}^{(\kappa)}\}$ (and $\{a_l^{(\kappa)}, q_{l}^{(\kappa)}\}$) for $l=1...\kappa$ of all evaluated RSB-orders $\kappa=1,2,...,200$. These data become dense for large enough $\kappa$ and approach the desired order function $q(a)$ in the limit $\kappa\rightarrow\infty$, which can be viewed as a fixed point function $q^*(a^*)$.

The unusual form $q^*(a^*)$ can be justified as follows: the parameters $a_l^{(\kappa)}$ and $q_l^{(\kappa)}$ can be viewed as functions of the continuous variable $l/\kappa\rightarrow\zeta$ in the $\kappa\rightarrow\infty$ limit. The fixed point solutions $a^*(\zeta)$ and $q^*(\zeta)$ can be combined by eliminating the variable $\zeta$, which results in the special form $q^*(a^*)$ where the variable itself is made up from continuously distributed fixed points. In the following we use $q(a)$ and $q^*(a^*)$ synonymously and distinguish them only if necessary.

\begin{figure}[here]
\resizebox{.5\textwidth}{!}{%
\includegraphics{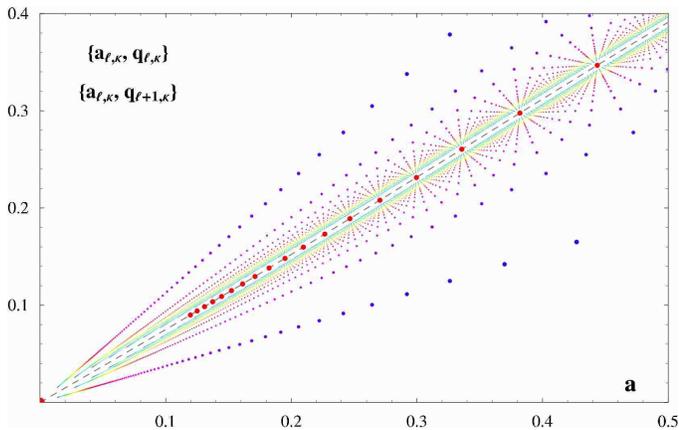}
}
\caption{\label{orderf-fixpoints}Self-consistent $\{a,q\}$-data in the small-$a$ regime are shown for all $\kappa=1,2,..200$ RSB orders, together with an analytical model function $q(a)$ (dashed grey line). Solutions $\{a_{l,\kappa},q_{l+1,\kappa}\}$ (above the line) and $\{a_{l,\kappa},q_{l,\kappa}\}$ (below) approach fixed points under constrained RSB-flow obeying $\kappa=\kappa(l)$, $l$ level number. Fixed point examples for $\kappa=l(m+1)/m-k/m-1$, $23\geq m\geq 6$ and $k$ integer are shown (red dots) including the fixed point $(0,0)$ (RSB-flow along $\kappa=l+k$). The analytical model function for $q(a)$ matches well all fixed points.
}
\end{figure}

\begin{figure}[here]
\resizebox{.5\textwidth}{!}{%
\includegraphics{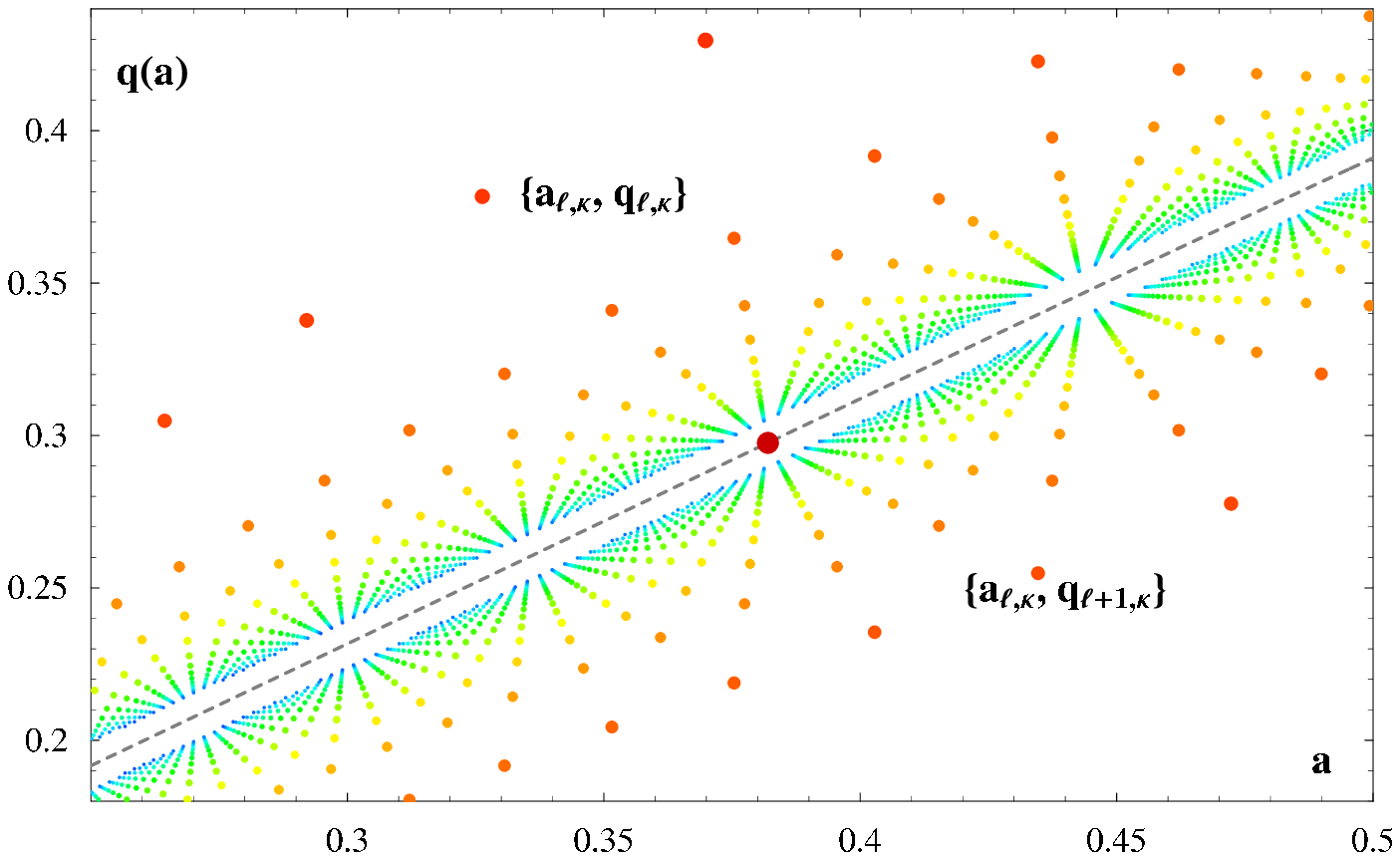}}
\resizebox{.5\textwidth}{!}{%
\includegraphics{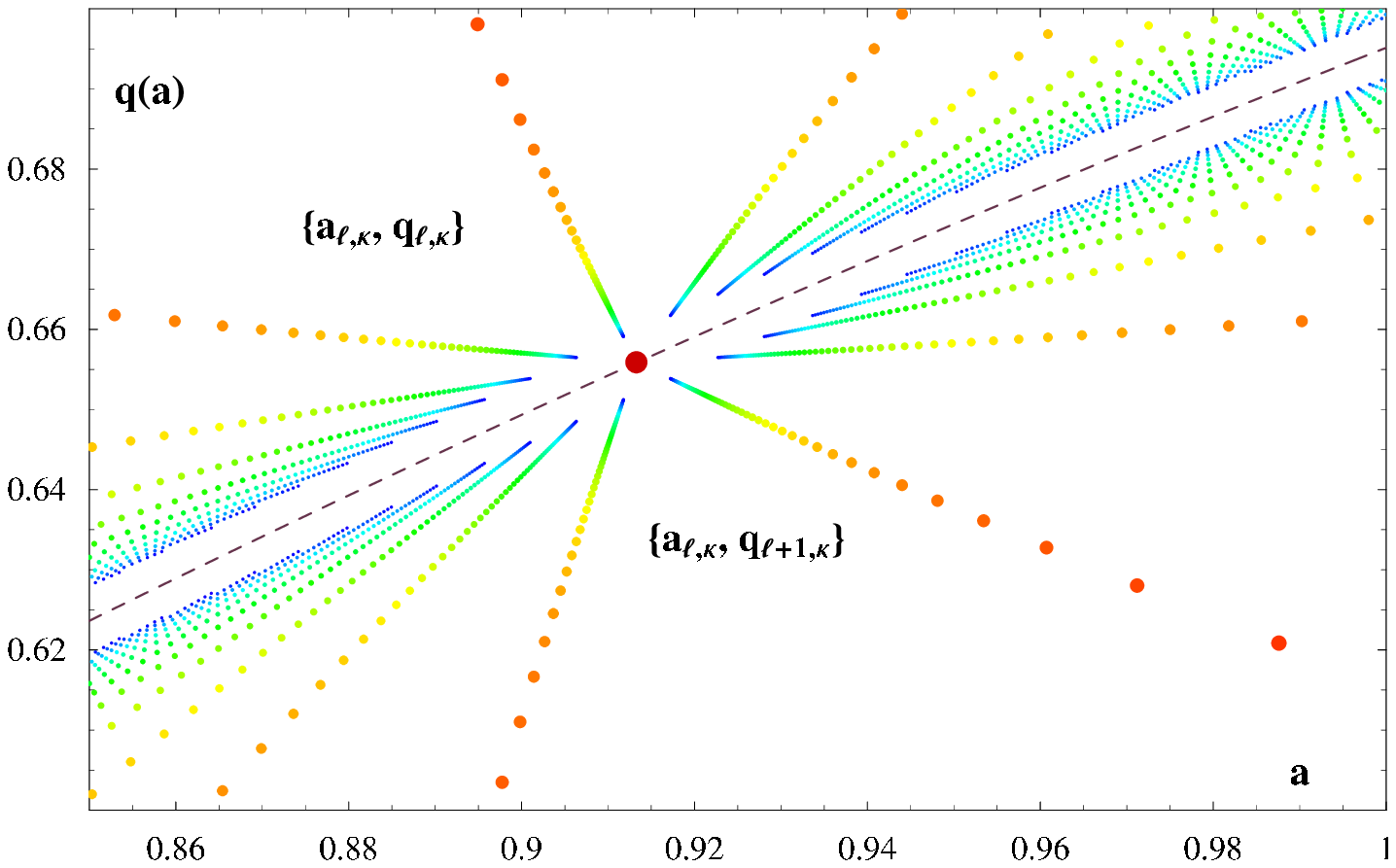}}
\caption{\label{fig:q(a)fixedpoint-vicinity}Two examples of alignment of solutions $\{a,q\}$ under constrained RSB-flow with $\kappa(l)$ (or $l(\kappa)$) towards two fixed points (big dark-red dots): (top) $p^*_{m=7,n=6}\equiv (a^*,q^*)_{m=7,n=6}=(0.3820, 0.2976)$  and (bottom) $p^*_{m=3,n=2}\equiv (a^*,q^*)_{3,2}=(0.9133, 0.6559)$. Pad\'e-line intersections, shown for $p^*_{7,6}$ in Fig.\ref{fig:fp038-padefit} and for a large-$a$ fixed point in Fig.\ref{fig:fp10p6-padefit}) determine the fixed points. Closer look shows that RSB fix-points lie dense and lead to a fixed point function, well approximated by the model function $q(a)$ (dashed grey).}
\end{figure}

At finite large orders one may define interpolating functions $q_{l+1}(a_l)\rightarrow q_<(a)$ and $q_{l}(a_l)\rightarrow q_{>}(a)$ which yield lower and upper bounds for the exact solution $q(a)\equiv q^*(a^*)$ at each value of $a$. Figure \ref{fig:q(a)-bounds} illustrates that this channel between lower and upper bound becomes extremely small for high orders $\kappa=O(10^2)$.
An illustration of the exact $q(a)$ being confined within such a channel as the RSB order $\kappa$ increases towards infinity, is provided in a more detailed way by zooming different regions of crossover between ${\cal CP}1$ and ${\cal CP}2$ in Figures \ref{orderf-fixpoints},\ref{fig:q(a)fixedpoint-vicinity}.

Figure \ref{fig:q(a)-bounds} moreover shows deviations of $q(1/a)$ from Gaussian behavior, which is a good approximation for small $1/a$.
%which would result if Pankov's $a=\infty$-expansion were extended to arbitrary values of $a$.
In both representations $q(a)$ and $q(1/a)$ it illustrates the existence of special lines which terminate obviously in fixed points - in fact there is a hierarchy of fixed points lying dense on the interval $0\leq a\leq\infty$. We shall make explicit use of these fixed points below.

Indeed, $200$ calculated orders of RSB for $T=0$ already yield an almost continuous function $q_{l}\left(\frac{a_l +a_{l+1}}{2}\right)$ which finally turns into $q(a)$ in the RSB-limit. Our previously published analytical model function \cite{prl2007} satisfies almost perfectly this constraint; as mentioned in Ref. \onlinecite{prl2007} it turned out that a small 'mass' function $w(a)$ in
\begin{equation}
q_{model}(a)=\frac{a}{\sqrt{a^2+w(a)}} {_1}F_1\left(\alpha,\gamma,-\frac{\xi^2}{a^2+w(a)}\right)
\label{model-function}
\end{equation}
models even the full crossover regime. The function $w(a)$ tends to a small constant $w(0)\approx 0.067$. This cutoff (of in general nonanalytic behavior for arbitrary parameters $\alpha$, $\gamma$) guarantees a strictly linear $q(a)$ relation in accordance with our high order data.
In the crossover regime between these two dynamic critical points, $w(a)$ can be modeled to depress the maximum error of $q(a)$ below $O(10^{-4})$ at each pseudo-time. A unique choice of $w(a)$ is not yet found, but excellent fits are obtained with $w(a)$ monotonically decreasing from $w(0)\approx 0.067$ to $w(\infty)=0$.
Using the high order data we have thus been able to improve the analytic approximation of the $T=0$ order function $q(a)$.
\begin{figure}[here]
\resizebox{.5\textwidth}{!}{%
\includegraphics{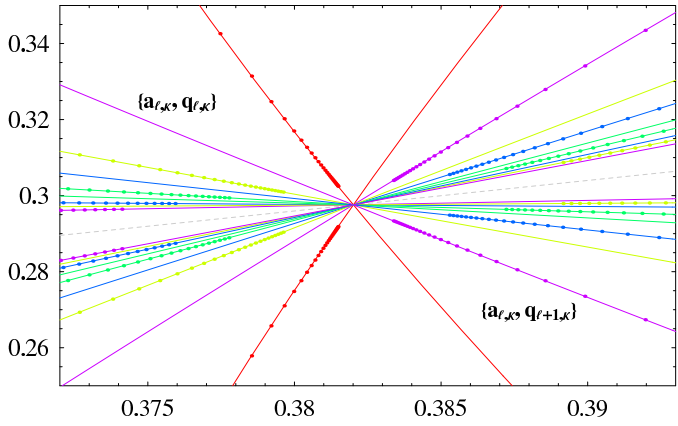}
}
\caption{\label{fig:fp038-padefit}Pad\'e approximants, matching the RSB-flow of discrete $q_{l,\kappa}(a_{l,\kappa})$-data ($1\leq\kappa\leq 200$) along $\kappa=\frac{7}{6}l-k/6-1$ for integers $k=-4,-3,...,4$, with initial values $\l_0=6+k,\kappa_0=6+k$ and $\Delta l=6$. All lines join in the RSB fixed point $p^*_{7,6}=(0.3820, 0.2976)$ (red dot) as $\kappa\rightarrow\infty$. (Pade curves are displayed here without termination at the fix point.Colors distinguish curves, eg red curves: $k=0$)}
\end{figure}
\begin{figure}[here]
\resizebox{.5\textwidth}{!}{%
\includegraphics{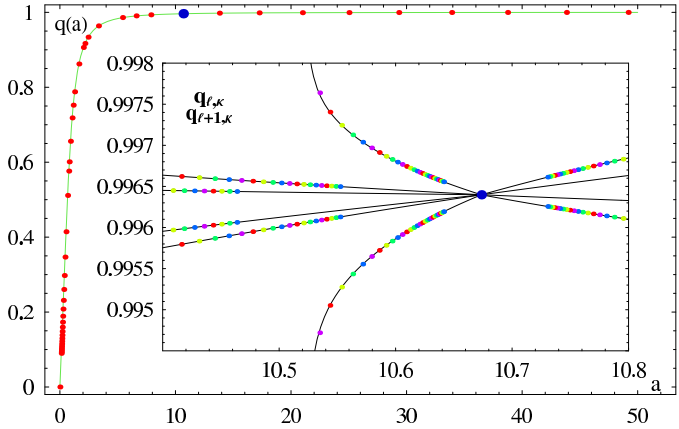}
}
\caption{\label{fig:fp10p6-padefit}Fixed point example in the $a\gg 1$-regime is displayed with nonlinear RSB-flow along $\kappa=6 l+k, k=-2,-1,..,2$, modeled by Pade-curves ending in a large-$a$ fixed point. Its position (big blue dot) at $\{10.6736,0.9964\}$ and that of a set of $50$ other fixed points (red dots), along the model function $q(a)$ (green line), is also shown.}
\end{figure}
%
%XXXXXXXXXXXXXXXXXXXXXXXXXXXXXXXXXXXXXXXXXXXXXXXXXXXXXXXXXXXXXXXXXXXXXXXXXXXXXXXXXXXXXXXXX
\subsection{Fixed points calculated from the RSB-flow towards $\kappa=\infty$}
%XXXXXXXXXXXXXXXXXXXXXXXXXXXXXXXXXXXXXXXXXXXXXXXXXXXXXXXXXXXXXXXXXXXXXXXXXXXXXXXXXXXXXXXXXX
%
The full set of self-consistent solutions for order parameters $q_l$ and (T-normalized) Parisi box sizes $a_l\equiv m_l(T)/T|_T=0$ can be described by matrix elements $p_{l,\kappa}\equiv \{a_{l,\kappa}\equiv a^{(\kappa)}_l,q_{l,\kappa}\equiv q^{(\kappa)}_l\}$ labeled by RSB-order $\kappa$ and level number $l$. Since the number of $q_l$-parameters exceeds by one the number of $a_l$-parameters (in each order of RSB), a second complementary set of matrix elements $\tilde{p}_{l,\kappa}\equiv \{a_{l,\kappa},q_{l+1,\kappa}\}$ should also be taken into account. These points $p_{l,\kappa}$ and $\tilde{p}_{l,\kappa}$ are displayed in the Figures $2-6$ and observed to approach the exact $q(a)\equiv q^*(a^*)$ along characteristic lines given below by Eq.(\ref{eq:k(l)-lines}) as $\kappa\rightarrow\infty$ ($p$ from above and $\tilde{p}$ form below $q(a)$ since $q_{l+1,\kappa}< q_{l,\kappa}$).

The set of all RSB-solutions up to a maximum order $\kappa$, as plotted in Fig.\ref{fig:q(a)-bounds} with a cutoff at $\kappa=200$, is then described by two triangular matrices with entries $\{a_{l,\kappa},q_{l,\kappa}\}$ (or with $\{a_{l,\kappa},q_{l+1,\kappa}\}$); the level-numbers $l$ run from $1$ to $\kappa$ for each RSB-order $\kappa$.

Along infinitely many lines in $(l,\kappa)$-space - the leading ones are very clearly visible in Figures \ref{orderf-fixpoints},\ref{fig:q(a)fixedpoint-vicinity} (and shown as calculated in Figs.5,6) - we observe very smooth behavior of slowly changing parameters $(a_{l,\kappa},q_{l,\kappa})$ which allow low order Pad\'e-approximants to match these data and to join in fixed points $p^*$ of the order function curve for $\kappa=\infty$. A special case is the origin where the fixed point is obtained with the extreme accuracy of $O(10^{-13})$).

Typical examples of such characteristic lines in $(l,\kappa)$-space can be given by the linear relation among the labels
\begin{equation}
\{l+k,\kappa=\frac{m}{n}l+k-1\}
\label{eq:k(l)-lines}
\end{equation}
(viewing $l\geq l_{min}\equiv l_0=n$ as the running index) with steps of $\Delta l=n$ and $m,n,k$ integer-valued.
The choice of $m/n$ selects one fixed point of the RSB-flow as $\kappa\rightarrow\infty$ with $l\rightarrow\infty$. Steps of $\Delta l=n$ are required to generate integer values for $\kappa$ (otherwise we wouldn't have numerical data). The integer $k$ distinguishes different lines which all meet in the same fixed point. Thus the fixed point $(a^*,q^*)$ is labeled by $m$ and $n$ or just by the rational number $m/n$. We have evaluated more than $50$ fixed points belonging to the exact order function $q(a)$. The higher $n$ the larger must be the steps $\Delta l$, hence one needs higher orders of RSB to find enough data points for reasonable curve-fitting through these points. This is one limitation of the method, but the almost linear character of a large number of these lines allows to calculate in principle a number of fixed points much larger than the order of RSB.
%
%XXXXXXXXXXXXXXXXXXXXXXXXXXXXXXXXXXXXXXXXXXXXXXXXXXXXXXXXXXXXXXXXXXXXXXXXXXXXXXXXXXXXXXXXX
\subsection{Discrete spectra in the $\kappa=\infty$ RSB-limit at zero temperature}
%XXXXXXXXXXXXXXXXXXXXXXXXXXXXXXXXXXXXXXXXXXXXXXXXXXXXXXXXXXXXXXXXXXXXXXXXXXXXXXXXXXXXXXXXXX
%
While the fixed point functions can be derived for all pseudo-time values $1/a$, the points $a=0$ and $a=\infty$ remain special limits. In a recent article \cite{prl2007} we have shown that infinitely large subclasses of certain self-consistent parameters ratios remain discrete at $T=0$ or $H=0$ even in the continuum limit. These discrete levels reside in the limits $a=0$ and $a=\infty$ when $\kappa=\infty$. Finite temperatures lift the discrete spectrum at $a=\infty$ into the continuum, while a magnetic field has a similar effect on the discrete levels at $a=0$. The ratios assume the value $1$ then. The discrete spectra therefore emphasize the critical nature of the points $a=0$ and $a=\infty$. We present in the following subsections new results for these $T=0$ levels of parameter ratios and, in Section \ref{finite-T-scaling}, describe their singular finite $T$ crossover.
%XXXXXXXXXXXXXXXXXXXXXXXXXXXXXXXXXXXXXXXXXXXXXXXXXXXXXXXXXXXXXXXXXXXXXX
\subsubsection{Level distribution at \cal{CP}2}
%XXXXXXXXXXXXXXXXXXXXXXXXXXXXXXXXXXXXXXXXXXXXXXXXXXXXXXXXXXXXXXXXXXXXXX
At the critical point  ${\cal CP}2$ the sub-class of small self-consistent parameters $q_k$ and $a_k$, which vanish in the $\infty$RSB limit (and condense into ${\cal CP}2$), obey
\begin{equation}
\frac{q_{\bar{l}+2}}{q_{\bar{l}+1}} = \frac{2l-1}{2l+1}
\quad {\rm and}\quad \frac{a_{\bar{l}+1}}{a_{\bar{l}}}=\frac{l}{l+1},
\end{equation}
with $\bar{l}\equiv \kappa-l$ and $l=1,2,...$; thus the ratios of these parameters are discrete and almost equidistant \cite{prl2007}.
Recurring these relations to the smallest parameters of each RSB-order $\kappa$, hence to $q_{\kappa+1}$ and $a_{\kappa}$ respectively, we obtain
\begin{equation}
%q_{\bar{k}+1}=(2k+1)q_{\kappa+1},\quad a_{\bar{k}}=(k+1)a_{\kappa}.
q_{\kappa+1-l}=(2l+1)q_{\kappa+1},\quad a_{\kappa-l}=(l+1)a_{\kappa}.
\end{equation}
The RSB flow of numerical data up to 200RSB allow to conclude that these minimal parameters vanish like
\begin{eqnarray}
q_{min}&\equiv& q_{\kappa+1} = \frac{1.03059}{\kappa}+\frac{1.31705}{\kappa^2}+O(1/\kappa^3),\\
a_{min}&\equiv& a_{\kappa} = \frac{2.77275}{\kappa}+\frac{3.54347}{\kappa^2}+O(1/\kappa^3).
\end{eqnarray}
The discretized slope of the order function in the point $a=0$, assumes the 200RSB value
\begin{equation}
\frac{q_{\bar{l}}-q_{\bar{l}-1}}{a_{\bar{l}}-a_{\bar{l}-1}}=\frac{2 q_{\kappa+1}}{a_{\kappa}}=0.74345
\end{equation}
or, by Pad\'e approximation of the RSB flow and extrapolation to $\infty-$RSB, one obtains
\begin{equation}
q'(0)=2 \hspace{.1cm} lim_{\kappa\rightarrow\infty} \frac{q_{\kappa+1}}{a_{\kappa}}=0.743368.
\end{equation}
As the calculation of fixed points of q(a) in the linear small-a regime shows, this agrees with the slope of the continuous $q(a)$ for $a\rightarrow 0$. The slope of the order function in ${\cal CP}_2$ provides one almost exact constraint for the order function
$$q'_{model}(a=0)=\frac{1}{\sqrt{w(0)}}{_1}F_1\left(\alpha,\gamma,-\frac{\xi^2}{w(0)}\right)=0.743368.$$
%
%XXXXXXXXXXXXXXXXXXXXXXXXXXXXXXXXXXXXXXXXXXXXXXXXXXXXXXXXXXXXXXXXXXXXXXXXXXXXX
\subsubsection{Level distribution at \cal{CP}1}
\begin{figure}[here]
\hspace{-.2cm}
\resizebox{.5\textwidth}{!}{%
\includegraphics{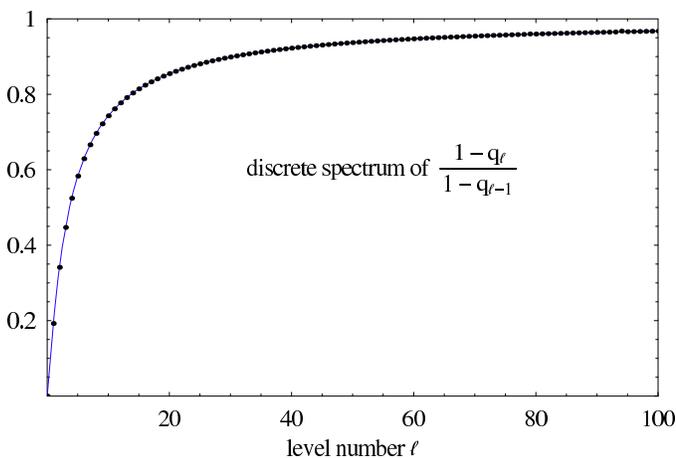}
}
\caption{\label{fig:large-q-ratios}The finite and discrete ratios of $1-q_l$-factors, which vanish for $\kappa\rightarrow\infty$ at $T=0$, as obtained by Pad\'e extrapolation to the RSB limit, plotted versus the level number $l$.
%(fit function with exponent $1.3$ does not represent a scaling law).
}
\end{figure}

\begin{figure}[here]
\hspace{-.5cm}
\resizebox{.5\textwidth}{!}{%
\includegraphics{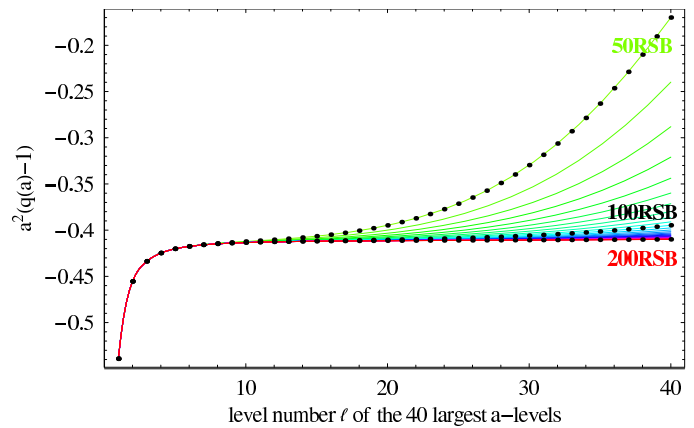}}
\caption{\label{fig:large-q-coefficient}The Figure shows convergence of the large-$a$ behavior of $a^2(q(a)-1)$ towards $\approx -0.41$. Some levels of the discrete spectrum at $a=\infty$ show a deviation.}
\end{figure}

%XXXXXXXXXXXXXXXXXXXXXXXXXXXXXXXXXXXXXXXXXXXXXXXXXXXXXXXXXXXXXXXXXXXXXXXXXXXXXX
In the large $a$ limit the characteristic feature are discrete spectra of $1-q_l$-ratios, which are shown in Figure \ref{fig:large-q-ratios}. In addition Figure \ref{fig:large-q-coefficient} shows that the $\frac{1}{a^2}$ coefficient of the almost continuous order function converges towards 0.41 except for the largest a-levels.
At zero temperature the order function differs from $1$ by $0.41/a^2$. Thus,
according to the large-$a$ expansion of our analytical model, the expansion coefficient is constrained to satisfy
\begin{eqnarray}
q(a)&=&1-\frac{\alpha\xi^2}{\gamma}\frac{1}{a^2}+O(1/a^{4})\nonumber\\
&=& 1-0.41\frac{1}{a^2}+O(1/a^4),
\end{eqnarray}
putting a constraint on $\alpha\xi^2\gamma$. Further constraints can be found from very precise numerical characteristics; it is planned to  use this analysis to narrow down the choice of an analytical order function model.

The discrete spectrum yields a coefficient which differs notably from this value, as one can see from Figure \ref{fig:large-q-ratios} (right) for the leading divergent $a_l$ parameters.

%
%
%XXXXXXXXXXXXXXXXXXXXXXXXX          II.1               XXXXXXXXXXXXXXXXXXXXXXXXXXXXXXXXXXXXXX
\subsection{Approach of equilibrium at $T=0$: leading and sub-leading scaling contributions}
%XXXXXXXXXXXXXXXXXXXXXXXXXXXXXXXXXXXXXXXXXXXXXXXXXXXXXXXXXXXXXXXXXXXXXXXXXXXXXXXXXXXXXXXXXXXX
%
The nonequilibrium susceptibility $\chi_1$ is a characteristic quantity measuring the distance from the equilibrium solution at $\kappa=\infty$. The entropy had been seen \cite{expcpaper} to vanish like the square of $\chi_1$.
The numerical solutions\cite{expcpaper} for $\chi_1$, evaluated for all $200$ leading RSB-orders, are well fitted by the $T=0$-form
\begin{eqnarray}
\label{eq:kappa-decay}
& &\chi_1(\kappa,T=0)\cong \frac{0.86}{(\kappa_0+\kappa)^{5/3}}+\frac{1.85}{(\kappa_0+\kappa)^4}+...\\
&\hspace{-.2cm}=& 0.86 \kappa^{-5/3}-1.83 \kappa^{-8/3}+3.12 \kappa^{-11/3}+1.85 \kappa^{-4}+...\nonumber
\end{eqnarray}
with $\kappa_0\cong 1.278$. As discussed in Ref.\onlinecite{expcpaper} the numerical uncertainty of $O(10^{-6})$ in the exponent is so small that the expectation of a rational-valued exponent due to one-dimensionality leads to the firm prediction of $\chi_1\sim\kappa^{-5/3}$. The quality and density of the numerical results is even high enough to predict the subleading correction and the amplitudes as well.
\begin{figure*}
\resizebox{.8\textwidth}{!}{%
\includegraphics{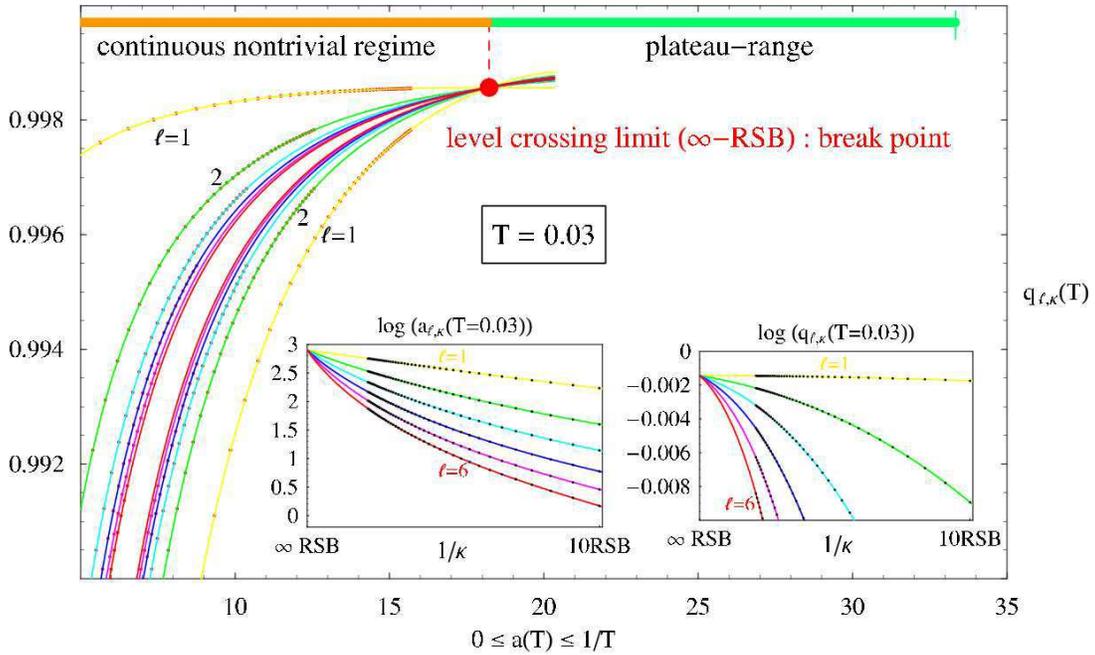}}
\caption{\label{fig:breakpoint}Main Figure shows the RSB-flow of $\{a_l(\kappa),q_l(\kappa)\}$ (above) and $\{a_l(\kappa),q_{l+1}(\kappa)\}$ (below), for each of the six highest parameter levels $(l=1...6)$ at fixed temperature $T=0.03$. Pad\'e approximations (shown for largest level (orange), 2nd-largest (green) etcetera) model the flow $\kappa\rightarrow\infty$ by extrapolation. $\kappa=\infty$ is reached in the level crossing point ${\cal LCP}=\{a_k(\kappa)\approxeq 18.226,q_k(\kappa)\approxeq 0.9986\}$ which separates plateau-regime ${\cal LCP}\leq a \leq 1/T$ from the $\infty$-RSB continuum $0\leq a\leq {\cal LCP}$. The Parisi box size $m(T=0.03)=0.5467$ is the break point value. Inserts illustrate that $q_l$- and $a_l$-levels meet in the ${\cal LCP}$ for $\kappa=\infty$.}
\end{figure*}
%
%XXXXXXXXXXXXXXXXXXXXXXXXXXXXXX             1          XXXXXXXXXXXXXXXXXXXXXXXXXXXXXXXXXXXXXXXXX
\section{Finite temperature scaling near the critical point ${\cal CP}1$ $(a=\infty,T=0,H=0)$}
\label{finite-T-scaling}
%XXXXXXXXXXXXXXXXXXXXXXXXXXXXXXXXXXXXXXXXXXXXXXXXXXXXXXXXXXXXXXXXXXXXXXXXXXXXXXXXXXXXXXXXXXXXXXX
Naturally one would like to start with a scaling hypothesis for the free energy $F$. However the SK-model has two critical points at $T=0$ and the free energy picks up contributions from both; in the RSB-limit, it can be expressed by integrals over entire crossover range from $a=0$ $({\cal CP}_2)$ to $a=\infty$ $({\cal CP}_1)$ involving the order function $q(a)$.

Thus it turns out useful to start with the scaling behavior of the self-consistent parameters $a_l$ and $q_l$, which teaches us how to embed scaling into the order function $q(a,T)$ mediating the crossover between the two critical regimes. Finally, by expressing free energy and internal energy in terms of the order function, and by linking the entropy with the non-equilibrium susceptibility, we shall arrive at consistent scaling predictions for $F$, $U$, and $S$ b+elow.

Let us begin with temperature-normalized block size parameters
\begin{equation}
a_l(\kappa,T)\equiv \frac{m_l(\kappa,T)}{T}
\end{equation}
where we consider first scaling in the $(\kappa,T)$-plane for fixed label $l$.
We must analyze the singular behavior near the critical point ${\cal CP}1$, where diverging $a_l(\kappa,T=0)\rightarrow\infty$ for $\kappa\rightarrow\infty$ lead to discretely spaced ratios $a_l(\infty,0)/a_{l-1}(\infty,0)$ in the $\infty-$RSB limit.
We identified the large order power law divergence
%-----------------------
\begin{equation}
a_l(\kappa,T=0)\sim \kappa^{5/3},\quad {\kappa\rightarrow\infty}
\label{eq:5/3-law}
\end{equation}
%-----------------------
for the subclass of large parameters $a_l$ (their number also grows to infinity as $\kappa\rightarrow\infty$).

The linear temperature decay of all Parisi box sizes $m_l(\kappa,T)\sim T$ holds for all {\it finite} RSB-orders, but not all $m's$ should vanish in the RSB limit at zero temperature, since the break point is not expected to be at $m_{1}=0$ (even in the $T\rightarrow 0$-limit\cite{Crisanti2002}). Thus, one should describe a non-commuting limits $T\rightarrow0$ and $\kappa\rightarrow\infty$ properly.

The Taylor series, valid as a low temperature expansion for any fixed finite RSB-order,
\begin{equation}
m_l(\kappa,T)\equiv a_l(\kappa,T)\hspace{.1cm}T=a_l(\kappa,0)T+\frac12 a_l'(\kappa,0)T^2+O(T^3)
\end{equation}
will anyway break down for those levels $l$ for which the expansion coefficients diverge as $\kappa\rightarrow\infty$. In accordance with the anomalous power law (\ref{eq:5/3-law}) it will be shown below by means of the fixed point order function that the correct scaling form for this ${\cal CP}_1$-divergent parameter sub-class reads
\begin{equation}
a_l(\kappa,T)=\kappa^{5/3} f_{a_l}(T/\kappa^{-5/3}),
\end{equation}
where the scaling function is well approximated by a low order $(2,3)$) Pad\'e series (one may also use $(1,2)$ or $(3,4)$ series)
\begin{equation}
f_{a_l}(x)=\frac{c_{0,l}+c_{1,l} x}{1+d_{1,l} x+d_{2,l} x^2}.
\end{equation}

This form fits well the available finite $T$ data up to $50$-RSB and satisfies
\begin{equation}
f_{a_l}(0)=c_{0,l} \quad{\rm finite\hspace{.1cm} and}\quad f_{a_l}(x)\sim \frac{1}{x}\quad {\rm for}\quad x\rightarrow\infty.
\end{equation}

The crossover line can be described by the characteristic (crossover) temperature
\begin{equation}
T_1(\kappa)\sim\kappa^{-5/3}.
\end{equation}

Beyond the crossover line, for temperatures $T\gg T_1(\kappa)$, the box sizes $m_l(x)=x\hspace{.1cm}f_{a_l}(x)$, which belong to the ${\cal CP}1$-divergent sub-class of $a_l$'s,
approach finite temperature-independent values. One obtains
\begin{equation}
\label{l-dep}
\lim_{x\rightarrow\infty}m_l(x)=c_{1,l}/d_{2,l}.
\end{equation}
While direct fits of our numerical data yield already a crude estimation of $m_1(\infty)$ for the break point, it was mentioned in Ref.\onlinecite{expcpaper} that $50$-RSB is not sufficient to determine the break point for arbitrary low temperatures. Yet, for $T=0.015$ a reliable break point value was determined by another procedure.

Here we are interested to obtain a good estimation of the breakpoint in close connection with the scaling picture. Therefore we employ the fixed point method and indeed succeed in finding a good approximation down to even lower temperatures and also answer the question whether the limit $m_l(\infty)$ in Eq.(\ref{l-dep}) shows a level index dependence or not. For finite $m_l$ an $l$-dependence would have implied a discrete distribution.
We shall find in subsection\ref{subsec:breakpoint} that all ratios become level-independent in the large $x$ limit
\begin{equation}
\frac{m_l(x)}{m_{l-1}(x)}=\frac{a_l(x)}{a_{l-1}(x)}=\frac{f_{a_l}(x)}{f_{a_{l-1}}(x)}\rightarrow 1\quad
{\rm for}\quad x\rightarrow\infty.
\end{equation}
The crossover from discrete parameter spectra for $T\ll T_1(\kappa)$ to the continuum on the other side of the crossover line, for $T\gg T_1(\kappa)$, is a rather singular effect mediated by the scaling function. We introduced above a scaling function which allows to suppress the discrete spacing between $q$- and $a$-parameters as one moves through the crossover line
$T_1(\kappa) \sim \kappa^{-5/3}.$
%
%XXXXXXXXXXXXXXXXXXXXXXXXXXXXXXXXXXXXXXXXXXXXXXXXXXXXXXXXXXXXXXXXXXXXXXXXXXXXXXXXXXXXXXXXX
\subsection{Forbidden level crossing at finite temperatures determines the break point}
\label{subsec:breakpoint}
%XXXXXXXXXXXXXXXXXXXXXXXXXXXXXXXXXXXXXXXXXXXXXXXXXXXXXXXXXXXXXXXXXXXXXXXXXXXXXXXXXXXXXXXXXX
%
We employ now the RSB-fixed-point technique to extract approximate values for the break point for rather low temperatures.

For this purpose, we consider fixed finite temperatures $T$ and fixed level numbers $l$ (down to lowest $T$ and $l$ small to catch the diverging-$a$ subclass near ${\cal CP}1$) and study the RSB-flow of the solutions $\{a_l(\kappa),q_l(\kappa)\}$, and also those of the complementary type $\{a_l(\kappa),q_{l+1}(\kappa)\}$, from low orders up to $\kappa=50$ as illustrated by Fig.\ref{fig:breakpoint} for an arbitrarily picked temperature $T=0.03$. Pad\'e-approximants fit the RSB-flow well and these extrapolated curves meet precisely in the same point. These cu+rves would cross each other, but then violate the reality condition of the self-consistent method beyond the level crossing point. We consider the level crossing point therefore as the limit of the nontrivial part of the order function, hence as the breakpoint.

The scenario remains the same for arbitrary fixed temperatures, only the extrapolation range increases with the level number and therefore becomes less accurate for smaller temperatures. Yet reliable solutions were obtained down to temperatures $T\approx 0.005$. The Figure Insert emphasizes the fact that the solutions indeed reach the level crossing point as $\kappa\rightarrow\infty$.

Approaching zero temperature and the RSB limit along the crossover line, $x$ fixed, with  $T_1(\kappa)\sim\kappa^{-5/3}$, leads to a discrete set of different Parisi box sizes $m_l(\kappa=\infty,T=0)$.
%
%
%XXXXXXXXXXXXXXXXXXXXXXXXXXXXXXXXXXXXXXXXXXXXXXXXXXXXXXXXXXXXXXXXXXXXXXXXXXXXXXXXXXXXXXXXX
\subsection{Nonequilibrium susceptibility $\chi_1$}
\label{subsec:chi1-scaling}
%XXXXXXXXXXXXXXXXXXXXXXXXXXXXXXXXXXXXXXXXXXXXXXXXXXXXXXXXXXXXXXXXXXXXXXXXXXXXXXXXXXXXXXXXXX
%
The scaling form of the non-equilibrium susceptibility $\chi_1(\kappa,T)$ can be given in terms of a scaling function $f_1$ by
\begin{equation}
\chi_1(\kappa,T)=\kappa^{-5/3}f_1(T/\kappa^{-5/3})
\end{equation}
where $f_1(x)\sim x$, $x\rightarrow\infty$, and $f_1(0)\approxeq 0.86$, reproduces the data and the leading $\kappa$-decay at $T=0$, as in Eq.(\ref{eq:kappa-decay}).
%XXXXXXXXXXXXXXXXXXXXXXXXXXXXXX        1     XXXXXXXXXXXXXXXXXXXXXXXXXXXXXXXXXXXXXXXXXXX
\section{Magnetic field scaling at critical point ${\cal CP}2$ (diverging pseudotimes $1/a\rightarrow\infty$)}
\label{finite-H-scaling}
%XXXXXXXXXXXXXXXXXXXXXXXXXXXXXXXXXXXXXXXXXXXXXXXXXXXXXXXXXXXXXXXXXXXXXXXXXXXXXXXXXXXXXXXXXXX
%
The magnetic field dependence at $T=0$ is expected to yield a plateau-like cutoff of the order function of similar shape as described in the Parisi form $q(x)$. We study now the field dependence of the smallest order parameter $q_{\kappa+1}(H,T=0)$ in $\kappa$-th order of RSB. $20$ orders of RSB turn out to be enough to extract the exponent describing the decay of $q_{\kappa+1}$ as the order of RSB tends to infinity. Guided by the results of finite temperature, where one single non-trivial rational exponent appeared, we observe an exponent $2/3$ to provide a reasonable picture for extrapolation towards $\infty$ RSB.

We first identify the $q_i\sim\kappa^{-1}$-law for (infinitely many) order parameters which vanish as $\kappa\rightarrow\infty$. The scaling hypothesis for $(\kappa,H)$-scaling, valid for the vanishing order parameters $q_i(\kappa,H)$, can be formulated as
\begin{equation}
q_i(\kappa,H,T=0)=\frac{1}{\kappa} f_{i}\left(\frac{H^{2/3}}{1/\kappa}\right)
\label{eq:Hfield-exponent}
\end{equation}
with $f_i(0)\neq 0$ and $f_i(x\rightarrow\infty)\sim x$.
\begin{figure}[here]
\resizebox{.5\textwidth}{!}{%
\hspace{-1.cm}
\includegraphics{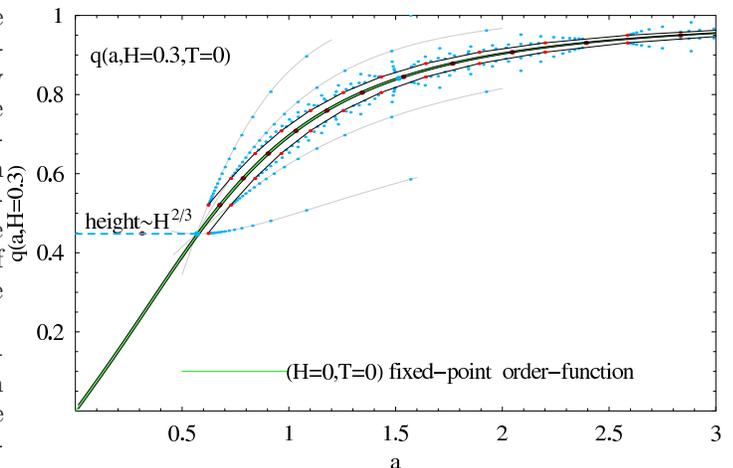}
}
\caption{\label{fig:q-a-cutoffH03}Upper and lower bounds of the order functions are shown for $\{20RSB, H=0.3\}$ and in 200RSB for zero field. For comparison the $H=0$ fixed point order function (green) is included. The discrete 20RSB (black dots) order (interpolation-)function and two fixed points obtained from the $1-20$ RSB-flow extrapolated to $\kappa=\infty$ (light-grey lines) in the finite field $H=0.3$ are displayed in addition. The fixed point at ${a^*_{pl}=0.568,q^*_{pl}=0.448}$ defines the plateau-cutoff of $q(a)$ in the finite field $H=0.3$.}
\end{figure}
\begin{figure}[here]
\resizebox{.5\textwidth}{!}{%
\hspace{-1.cm}
\includegraphics{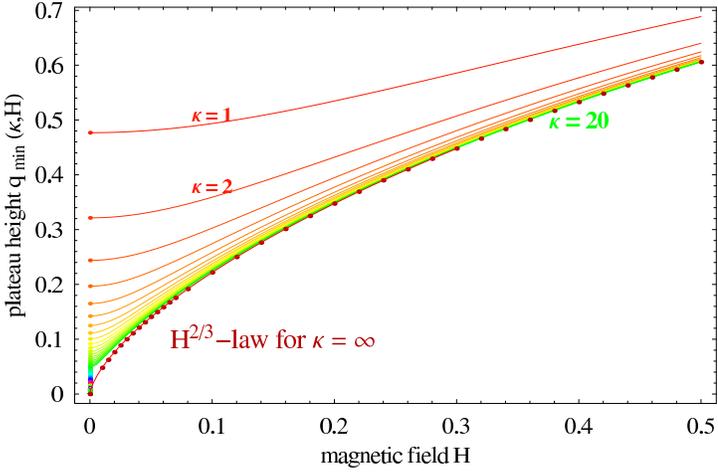}
}
\caption{\label{fig:field-exponent}Numerical data for the magnetic field range $0<H\leq 0.5$ are shown from 1st to 20th order of RSB (green). Dots (dark red) show the plateau-height small-q cutoff obtained from the fixed point order function. At $H=0$ dots show the calculated orders up to 200RSB for comparison. The RSB flow for $q_{min}(\kappa,H=0)$ is given by a $1/\kappa$ law, while $q_{min}(\kappa=\infty,H)$ obeys an $H^{2/3}$-law.}
\end{figure}
The numerical procedure chosen in order to arrive at this proposal has been to extrapolate to $\infty$RSB the smallest $q_1$-values at fixed non-vanishing small magnetic fields. The higher the field the less orders of RSB are needed (similar as in the case of finite temperatures). Twenty steps of RSB generate almost exact results down to $H\approx 0.15$. Extrapolation of the RSB-flow is hence reliable down to much smaller field-values, where one has already entered the critical regime. Thus many RSB fixed point values (at $\kappa=\infty$) are well approximated and can be used to match a power law w.r.t. the magnetic field. In this way the magnetic field exponent of Eq.(\ref{eq:Hfield-exponent}) is found to differ only by $0.003$ from the value $2/3$ which led to the assumption that this rational number is exact.
%XXXXXXXXXXXXXXXXXXXXXXXXXXXXXXXXXXXXXXXXXXXXXXXXXXXXXXXXXXXXXXXXXXXXXXXXXXXXXXXXX
\section{Scaling behavior of the free energy $F$, internal energy $U$, and entropy $S$}
\label{free-energy-scaling}
%XXXXXXXXXXXXXXXXXXXXXXXXXXXXXXXXXXXXXXXXXXXXXXXXXXXXXXXXXXXXXXXXXXXXXXXXXXXXXXXXXX
%
Low temperature expansions of internal energy $U$, entropy $S$, and the free energy $F=U+TS$ were reported in the framework of our high order RSB analysis, and found in agreement with already known results. In the present context of scaling theory, we also look for scaling of RSB-parameters together with temperature and also small field variation.
A useful way to study the RSB-flow in terms of $\kappa-$scaling is by invoking the internal energy formula at $T=0$ and $H=0$
\begin{eqnarray}
\label{eq:U}
& &\hspace{-.4cm}U(\kappa,T=H=0)=-\chi_1-\frac12 \sum_{l=1}^{\kappa}a_l (q_l^2-q_{l+1}^2) \\
& \hspace{-.4cm}\Rightarrow& \lim_{\kappa\rightarrow\infty}U(\kappa,0)=-\frac12 \int_{0}^{\infty}da\hspace{.1cm}(1-q(a)^2)
\end{eqnarray}
The summation includes contributions from both critical points and from the crossover regime in between. Consequently one cannot expect to obtain scaling laws from a single hypothesis imposed on the total free energy. The problem has more in common with critical dynamics, however with two critical points in the long pseudo-time limit ($1/a\rightarrow\infty$) and in the short pseudo-time limit ($1/a\rightarrow0$).
%XXXXXXXXXXXXXXXXXXXXXXXXXXXXXXXXXXXXXXXXXXXXXXXXXXXXXXXXXXXXXXXXXXXXXXXXXXXXXXXXXXXXXXXX
%\subsection{
%{\it Scaling contributions to free energy $F(\kappa,T,H)$ from the critical points ${\cal CP}1$ and ${\cal %CP}2$.}\\
%XXXXXXXXXXXXXXXXXXXXXXXXXXXXXXXXXXXXXXXXXXXXXXXXXXXXXXXXXXXXXXXXXXXXXXXXXXXXXXXXXXXXXXXX

As reported in Ref.\onlinecite{expcpaper} the free energy has a low T expansion in the RSB limit given by
$F=F(T=0)-S(T=0)T+\sum_{k=2} f_k T^k$, where the leading temperature behavior is $F(\kappa=\infty,T)-F(\infty,0)\sim T^3$.
The leading large-$\kappa$ correction of the $T=0$ free energy has also been reported to decay like $\kappa^{-4}$.

In the large-$a$ regime, temperatures scale like $\kappa^{-5/3}$ and hence the large-$a$ scaling contribution is $\delta F\sim \kappa^{-5}$. Thus the leading temperature dependence belongs to a sub-leading $\delta F\sim \kappa^{-5}$ correction.

We attempt to distinguish singular scaling contributions from both critical points from the non-singular contributions to the free energy.
The small $a$-regime contribution can be estimated from
\begin{eqnarray}
\label{eq:F}
& &F(\kappa=\infty,T=0)=U(\kappa=\infty,T=0)=E_0(H)\nonumber\\
& &=-\frac12 \int_0^{\infty}da\hspace{.1cm}(1-q^2(a))-M(H) H,
\label{eq:T0energy}
\end{eqnarray}
where $M(H)$ denotes the field-generated magnetization. Recalling the small-$a$ expansion of the order function, $q(a)\sim a-const. a^3+O(a^5)$, one must expect an $H^{10/3}$-contribution from the plateau-regime, which implies also an $O(\kappa^{-5})$ contribution.
The free energy data are compatible with an $H^{10/3}$ small field scaling part.

It must be concluded that the leading correction $\kappa^{-4}$ must originate in the intermediate $a$- regime (not yet identified in detail). It can, after all what was said before, not be assumed to be a scaling contribution. We should therefore attribute it to the regular free energy part.

The entropy was found to obey\cite{expcpaper}
$$S(\kappa,T=H=0)=-\frac14 \chi_1(\kappa,T=H=0)^2.$$
It is known that only the large-a regime near ${\cal CP}1$ is responsible for the leading $\kappa$-behavior of $\chi_1$ at zero temperature, hence this holds also for the $T=0$-entropy. Since thermal behavior is also caused by the ${\cal CP}1$ contributions, we can therefore
claim that the scaling-contribution to the entropy obeys
\begin{equation}
S_s(\kappa,T)=\kappa^{-10/3}f_S(T^2/\kappa^{-10/3}),
\end{equation}
with $f_S(x)=-0.72\hspace{.1cm}x$ (see Ref.\onlinecite{expcpaper}) for $x\rightarrow\infty$ and
$f_S(x)\approxeq -0.185$ for $x\rightarrow0$.
Thus the entropy contributes to the leading $O(T^3)$ low temperature correction of the free energy.
This $TS$-term contributes again only a sub-leading correction $\delta F\sim \kappa^{-5}$ from the large $\kappa$-scaling regime.

Let us recall the large-$\kappa$ dependence of the free energy at zero temperature, well described by the optimal fitting form
$$F(\kappa,T=0)=F(\infty,0)+\frac{c_4}{(\kappa+\kappa_0)^4}+\frac{c_5}{(\kappa+\kappa_0)^5}+..,$$
where excellent Pad\'e-fits yield the constant $\kappa_0=1.28$.
The leading correction $\kappa^{-4}$ does neither originate from the scaling regime near ${\cal CP}1$ nor from that near ${\cal CP}2$, and hence must be expected not to scale. We therefore consider it as part of a regular $F$-contribution $F_{reg}(\kappa,T,H)$.

%$=F(\infty,0)-\chi(\infty,0) H^2$ in zero field.
Thus we propose that the free energy consists of a sum of a regular and of two singular parts, where the latter ones scale according to whether they are ${\cal CP}1$- or ${\cal CP}2$-critical.

As a consequence of this two-critical point picture and in agreement with the numerical data,
we separate two singular contributions, which offer different scaling behavior, from a regular part $F_{reg}$ by
\begin{equation}
F(\kappa,T,H)=F_{reg}(\kappa,H,T)+F^{({\cal CP}1)}_s(\kappa,T)+F^{({\cal CP}2)}_s(\kappa,H)
\end{equation}
where the magnetic-field controlled critical point ${\cal CP}2$ and the temperature-controlled critical point ${\cal CP}1$ contribute respectively
\begin{equation}
F^{({\cal CP}1)}_s(\kappa,T)=\kappa^{-5}f_{cp1}(T/\kappa^{-5/3}),
\end{equation}
and
\begin{equation}
F^{({\cal CP}2)}_s(\kappa,H)=\kappa^{-5}f_{cp2}(H^{2/3}/\kappa^{-1})
\end{equation}
with $f_{cp1}(x)\sim x^3, f_{cp2}\sim x^5$ for $x\rightarrow \infty$ and both finite for $x\rightarrow0$. This claim refers to the leading scaling behavior at ${\cal CP}1$ and ${\cal CP}2$; corrections with analytic $T$-dependence near ${\cal CP}2$ and analytic field-dependent corrections near ${\cal CP}1$ may occur.

A contribution $-\frac12 \chi(\kappa) H^2$-term, which yields the linear equilibrium susceptibility from $-\partial_H^2 F$, belongs to the regular part $F_{reg}$ with $\chi(\kappa\rightarrow\infty,T<T_c)=1$.

It is interesting trying to translate the given power laws into scaling with the number $N$ of spins for the finite $N$ SK-models
\footnote{We thank Thomas Garel for drawing our attention to the paper by Bouchaud et al\cite{bouchaud-energy-exponents}}
which corresponds to a finite size system with $N=L^d$, $d$ denoting the real space dimension.
Scaling with $L$ or $N$ delivered a leading correction $\sim N^{-2/3}$ for the finite SK-model \cite{boettcher,bouchaud-energy-exponents}.
If we would assume scaling of the leading correction $\kappa^{-4}$ with $N$, a scaling function depending on $N^{-1/6}/\kappa^{-1}$ would result\cite{thg-privcom,ds-privcom}. However this rests on the assumption that the leading $N^{-2/3}$ energy correction arises from the entire $a$-regime. Many open questions seem to show up here.
%
%XXXXXXXXXXXXXXXXXXXXXXXXXXXXXXXXXXXXXXXXXXXXXXXXXXXXXXXXXXXXXXXXXXXXXXXXXXXXXXXXX
\section{Fixed point distributions}
\label{U-distribution}
%XXXXXXXXXXXXXXXXXXXXXXXXXXXXXXXXXXXXXXXXXXXXXXXXXXXXXXXXXXXXXXXXXXXXXXXXXXXXXXXXX
%
%XXXXXXXXXXXXXXXXXXXXXXXXXXXXXXXXXXXXXXXX
\subsection{Ground state energy $E_0$}
\begin{figure}[here]
\vspace{.0cm}
\resizebox{.45\textwidth}{!}{%
%\hspace{-1.cm}
%\includegraphics{E0-distribution-a-int.eps}
\includegraphics{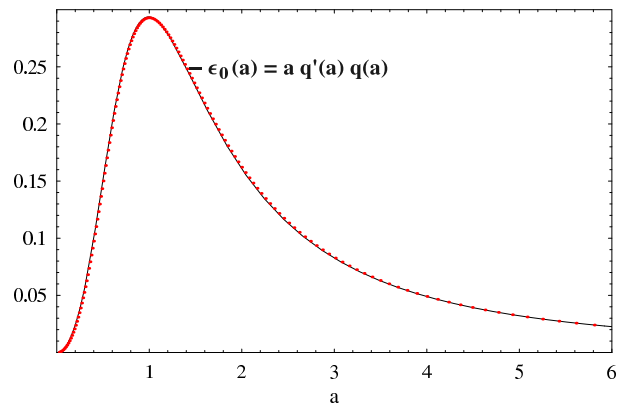}
}
\caption{\label{fig:energy-distrib1}The (negative) energy density $-\epsilon_{0}(a)$ of the ground state energy $E_0=\int_0^{\infty}da\hspace{.1cm}\epsilon_{0}(a)$, obtained from our analytical model function $q(a)$ as $-\epsilon(a)=a\hspace{.1cm}q'(a)q(a)$ (black underlying curve), is shown to agree well with discrete numerical $200$-RSB results
%(from $100$ (light blue) to $200$-RSB (red) in steps of $\Delta\kappa=10$)
.}
\end{figure}

%XXXXXXXXXXXXXXXXXXXXXXXXXXXXXXXXXXXXXXXX
We can extract more detailed information from our numerical analysis of RSB in the SK-model \cite{prl2007,expcpaper} beyond the calculation of the global ground state energy. The RSB-flow of the energy level distribution and naturally the energy density $\epsilon_0(a)$ as a function of pseudo-times can be given. In the latter case, a test of our analytic order function model against the numerical results \cite{expcpaper} is provided by the use of $q(a)$ and of $q'(a)$. Both are required in the ground state energy formula in Eq.\ref{eq:T0energy} according to
\begin{equation}
\label{gs-energy}
E_0=\int_0^{\infty}da\hspace{.1cm}\epsilon_0(a)=
-\int_0^{\infty}da\hspace{.1cm}a\hspace{.06cm}q'(a) q(a).
\end{equation}
Using the analytic form (3) and high RSB-order results for $\kappa=100,110,120,...200$, we obtain Fig.\ref{fig:energy-distrib1}.
\footnote{On the given scale, all numerical results fall almost exactly onto the single analytical curve for $\epsilon_0(a)$; only extreme magnification reveals the RSB-flow of the numerical data and tiny deviations from the analytical model in the crossover regime between ${\cal CP}1$ and ${\cal CP}2$.}
We do not find exponential tails in this energy distribution, instead observe simple power law decay in the limits of small and large $a$.

A second important representation shows the energy level contributions from
\begin{equation}
\quad \epsilon_0(l,\kappa)= -\frac14 \lim_{T\rightarrow 0} \hspace{.1cm} a_l(\kappa)\left(q_l^2(\kappa)-q_{l+1}^2(\kappa)\right),
\end{equation}
as a function of normalized level index $l/\kappa$, and with boundary conditions $a_0=\beta,\hspace{.1cm}q_0=1$.
The sum over all energy levels $\epsilon_0(l,\kappa)$ with level index $l=0,1,2,...\kappa$ for each calculated RSB-order yields the RSB-flow of the ground state energy
\begin{equation}
E_0(\kappa)=\sum_{l=0}^{\kappa}\epsilon_0(l,\kappa)
\end{equation}
towards the exact value\cite{expcpaper} $E_0(\kappa=\infty)=E_0$.
\begin{figure}[here]
\vspace{.0cm}
\resizebox{.5\textwidth}{!}{%
\hspace{-.2cm}
\includegraphics{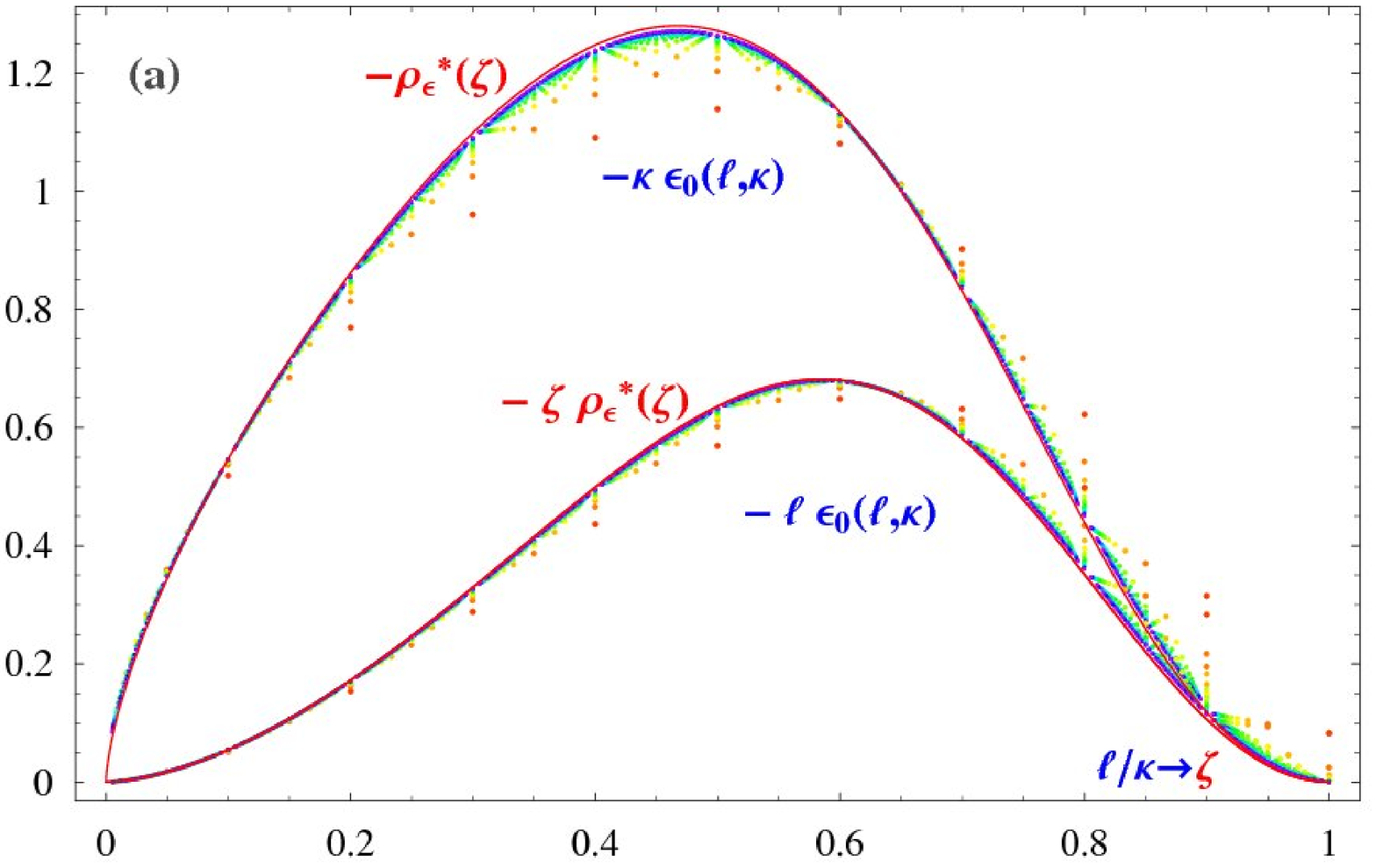}}
\resizebox{.45\textwidth}{!}{%
\hspace{-.4cm}
\includegraphics{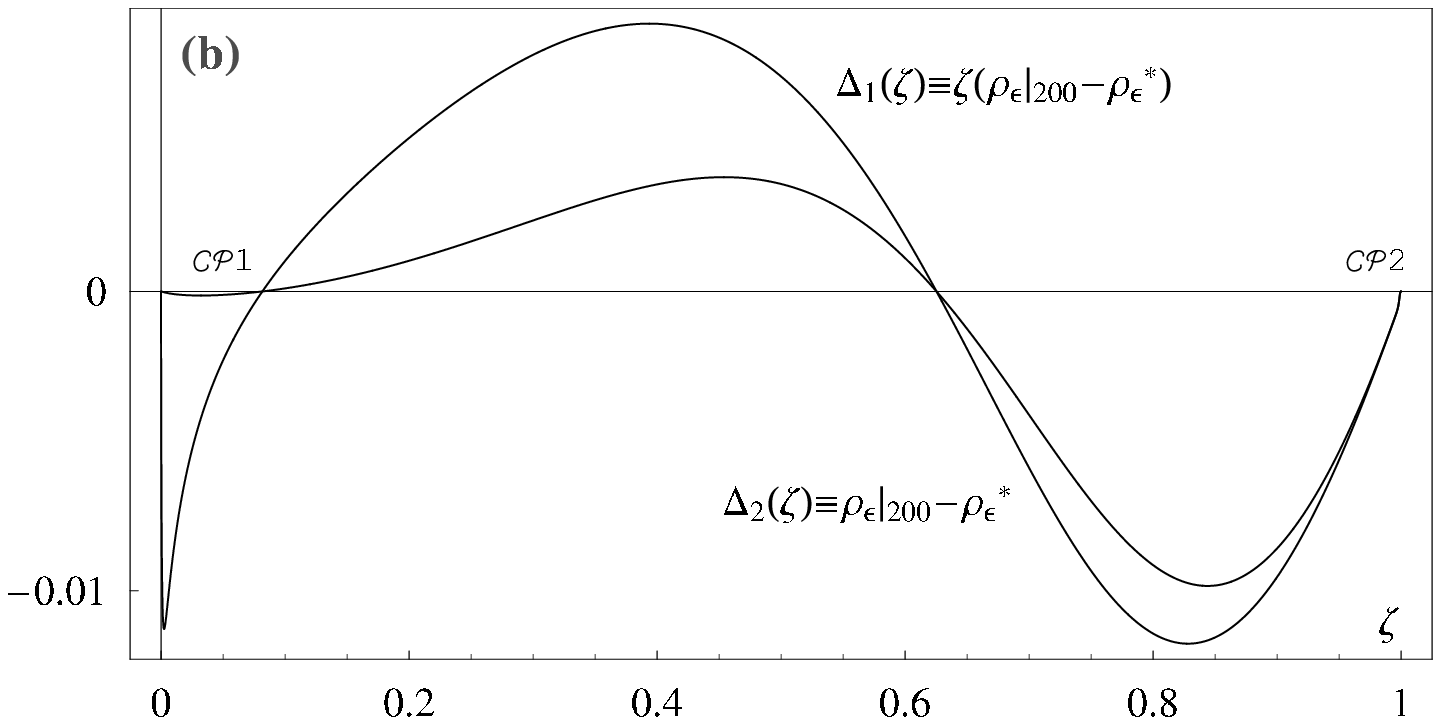}}
\caption{\label{fig:energy-distrib2}Part (a) shows ground state energy solutions (dots) for $\kappa\hspace{.1cm}\epsilon_0(l,\kappa)$ (or $l\hspace{.1cm}\epsilon_0(l,\kappa)$) and their RSB-flow from $\kappa=10$ to $\kappa=200$, $\Delta\kappa=10$, towards their fixed point functions $[\zeta]\rho_{\epsilon}^*(\zeta)$ on the unit interval of levels $l/\kappa\rightarrow\zeta$.
(b) shows differences $\Delta_{1,2}$ between 200RSB (interpolation functions $[\zeta] \rho_{\epsilon}|_{200}$) and the fixed point functions $[\zeta] \rho^*_{\epsilon}(\zeta)$ (red curves in (a)).}
\end{figure}
A proper normalization of level numbers by the RSB-order $\kappa$, displays the level-distributions for each RSB-order on the same interval of unit length. Subsequent rescaling of the energy level allows to visualize the RSB-flow towards one fixed point energy distribution (which of course depends on the rescaling factor
\footnote{One may choose rescaling factors such that discrete spacing of energy levels would survive even in the fixed point function ($\kappa=\infty$) near $l/\kappa=0$ and $l/\kappa=1$; this would correspond to the discrete spectra of parameter ratios discussed in the paper.}). Fig.\ref{fig:energy-distrib2} shows two choices ($l$- and $\kappa$-rescaling of $\epsilon_0(l,\kappa)$) - in both cases the convergence towards the fixed point function is obvious.

Fixed points (under RSB-flow) have been calculated in the same way as shown before for the order function. For example, fixing $l/\kappa$ to a rational number $m/n$ within the unit interval, one can see many of the leading fixed points in Fig.\ref{fig:energy-distrib2} following the RSB-flow along vertical lines fixed by $m/n$. The piecewise dense set of calculated fixed points was obtained by an extrapolated Pad\'e approximation for $n=2,...,51$ with $m=1,...,n-1$.
These fixed points are shown in Fig.\ref{fig:energy-distrib2} together with their fit function, obtained here as an $(8,8)$-Pad\'e series. The fixed points are piecewise dense with some gaps near 'leading' fixed points (eg at , which become however closed as one higher orders are evaluated.
The fit function (interpolating between the dense regions) represents an approximation for the exact fixed point energy distribution function $\epsilon_0^*(\zeta)$ with $l/\kappa\rightarrow\zeta$ in the $\infty$-RSB limit. The numerical integration of the approximated function $\rho_{\epsilon}^*(\zeta)$ (which corresponds to $\epsilon^*(a)$ of Eq.(\ref{gs-energy}) transformed form $0\leq a\leq\infty$ onto the unit interval $0\leq \zeta\leq1$) yields
\footnote{We tacitly assume here that the the density functions $\rho^*_{\epsilon}(\zeta)$ and also $\rho^*_{\chi}(\zeta)$ (below) are Riemann-integrable. The upgrade from the set of rationable numbers $l/\kappa$ to a continuous variable $\zeta$ could in principle hide a mathematically subtle problem, if the density functions were highly discontinuous and would for example require a Lebesgue integral}
\begin{equation}
E^*_0\equiv E^*(T=0)=\int_0^{1}d\zeta\hspace{.1cm}\rho_{\epsilon}^*(\zeta)|_{approx}\approx -0.76314.
\end{equation}
By reproducing the correct value\cite{expcpaper} up to $O(10^{-5})$, this provides a good test of the fixed point method. An alternative calculation, using Eq.(\ref{gs-energy}) with plugged in fixed point order function confirms the numerical value $E_0^*$.
The inserted figure shows the magnitude of energy-corrections per level $l$ occurring from $200$-RSB to the exact $\infty$-RSB energy per level (recall that $l$ labels the Parisi boxes of the RSB order parameter).

Different power law decays are observed in the small $l/\kappa$ (${\cal CP}1$) and in the $l/\kappa\approx 1$ range near (${\cal CP}2$).

An analytical modeling of the fixed point energy distribution must be attempted in the future; it might reveal more valuable information about the relation with directed polymers and/or with the KPZ-universality classes\cite{praehofer}.

Energy distribution functions play an important role in the characterization of directed polymers \cite{monthus-pre69,monthus-pre73,monthus-pre74}.
Generalized Gumbel statistics\cite{bertin-gumbel} were found to describe the statistical fluctuations of global quantities (like the energy). It is perhaps in this respect where a clear distinction between the directed polymers and the present universality class can be made. But this detailed comparison is beyond the scope of the present paper and should be attempted in the future.
%
%XXXXXXXXXXXXXXXXXXXXXXXXXXXXXXXXXXXXXXXXXXXXXXXXXXXXXXXXXXXXXXXXXXXXXXXXXXXXXXXXXXXXXXXXXXXXX
\subsection{Equilibrium susceptibility per level}
%XXXXXXXXXXXXXXXXXXXXXXXXXXXXXXXXXXXXXXXXXXXXXXXXXXXXXXXXXXXXXXXXXXXXXXXXXXXXXXXXXXXXXXXXXXXXX
To conclude this section we extend the described method to the $\chi(a)$-density of the equilibrium susceptibility $\chi$ and in particular to the distribution per level $l$.
In the RSB-limit, the total $\chi$ is known to be equal to $1$ in the entire ordered phase.
The RSB flow thus moves towards a fixed point function $\chi(a)=a\hspace{.1cm}q'(a)$ with the property $\int_0^{\infty} da \chi(a)=1$ (this had been used before as a constraint for our analytical order function model \cite{prl2005,prl2007}).

\begin{figure}[here]
\resizebox{.48\textwidth}{!}{%
\hspace{-.2cm}
\includegraphics{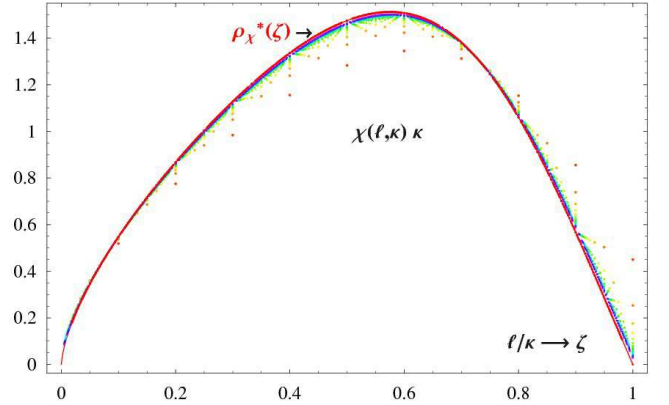}}
\caption{\label{fig:susc-density}RSB-flow of the equilibrium susceptibility per level $l$ from $\kappa=10$- to $\kappa=200$ (in steps of $10$ orders of RSB and rescaled by $l\rightarrow l/\kappa$ to a unit interval) towards the fixed point susceptibility distribution $\rho^*_{\chi}(\zeta)$ (black curve)}
\label{fig:chi}
\end{figure}

Let us now study the RSB-flow of the discrete representation $\chi(l,\kappa)=a_l(\kappa)(q_l(\kappa)-q_{l+1}(\kappa))$. The result for the susceptibility per level $l$ (normalized by RSB-order $\kappa$) is shown in Fig.\ref{fig:chi} (and corresponds to the energy per level distribution shown in the preceding figure).
The shape recalls universal distributions observed for the KPZ growth processes \cite{praehofer}. This relation or mapping must be studied in the future, particularly because - as explained in Ref.\onlinecite{praehofer} - the related statistical fluctuations have been associated with universal critical behavior.

Beyond the flow of the finite RSB orders $\kappa=10,20,30,...,200$ we have added the fixed point function $\rho^*_{\chi}(\zeta)$ for the susceptibility density (ie $\chi(a)$ transformed onto the unit interval $0\leq\zeta\leq1$) which must obey $\int_0^1 d\zeta\hspace{.1cm}\rho^*_{\chi}(\zeta)=1$.
A simple approximate calculation of the interpolating fixed point function reproduces the exact constraint with an error of only $O(10^{-5})$. Again this confirms the power of the method, which can eg be used to test analytical proposals.

Small changes from $200$-RSB to $\infty$-RSB are resolved in Fig.\ref{fig:susc-density} and in Fig.\ref{fig:energy-distrib2}. Their tendency is to make the distribution more symmetric. Yet the distribution per normalized level remains asymmetric as for the energy distribution (as a function of the dense levels $l/\kappa$), a universal fact that has been observed as a special feature of the SK-model in contrast to symmetrical distributions finite-range spin glasses.
%
%XXXXXXXXXXXXXXXXXXXXXXXXXXXXXXXXXXXXXXXXXXXXXXXXXXXXXXXXXXXXXXXXXXXXXXXXXXXXXXXXX
\section{Scaling with the pseudo-dynamical variable of the order function $q(a)$}
\label{pseudo-dynamical-scaling}
%XXXXXXXXXXXXXXXXXXXXXXXXXXXXXXXXXXXXXXXXXXXXXXXXXXXXXXXXXXXXXXXXXXXXXXXXXXXXXXXXX
%
In previous publications we found a Langevin-type representation \cite{prl2007,pssc2007} for a logarithmic derivative of the order function $q(a)$ with respect to $1/a$. This ordinary differential equation (without stochastic field) is much simpler than the exact partial differential equations, which is a consequence of the existence of scaling behavior and of homogeneous functions. It is well-known that scale invariance and the so-called similarity method reduces partial to ordinary differential equations \cite{debnath-book}. Therefore, at least near the critical points one can expect ordinary differential equations to describe RSB.

The Langevin-type of differential equation could however be reshaped in terms of different pseudo-dynamic variable $a$, $1/a$ or other forms. The differential equation remains to be relaxational and thus there remains some arbitrariness in the choice of the proper 'time' variable $\tau$.
If we wish to apply dynamic scaling to the RSB-representation, we are unfortunately bound to make a definite choice.
Let us consider $a+1/a$ as a pseudo-time in order to conform with the expectation that critical behavior at either of the points ${\cal CP}1$ or ${\cal CP}2$ should occur in the long-time limit. Then at ${\cal CP}1$ we would get $\tau\sim a\rightarrow\infty$ while $\tau\sim\frac{1}{a}\rightarrow\infty$ at ${\cal CP}2$.

We may now consider pseudo-dynamical scaling by studying the $a$-dependent quantities like the order function near ${\cal CP}1$ and ${\cal CP}2$.

Near ${\cal CP}1$ the order function obeys
\begin{equation}
q(a,\kappa,T)=1+a^{-2} f_q(T^2/a^{-2},a^2/\kappa^{10/3})
\end{equation}
with $f_q(x,0)\sim x$, $f_q(0,x)\sim x$, and $f_q(0,0)$ finite. In terms of the transformed order function $\phi\sim \partial_{1/a}log(q(a))$ one gets $\phi\sim 1/a$ at $\kappa=\infty, T=0$ and $\phi\sim T$ at $a=\infty,\kappa=\infty$. This would allow to extract an exponent $\beta=1$, and together with $\xi_{\kappa}\sim a^{3/5}\sim T^{-3/5}$ the correlation exponent $\nu=3/5$ results. Then, using $\tau\sim 1/a$ as a pseudo-time variable near ${\cal CP}1$, the dynamic exponent follows from $\tau\sim\xi_{\kappa}^z$ as $z=5/3$, remarking also that $z\hspace{.1cm}\nu=1$.

Given the already mentioned similarities with directed polymers, the known relationship between those and the KPZ-equation \cite{laessig-KPZ} suggests a comparison between pseudo-dynamics of RSB in the SK-model and the dynamic KPZ-behavior.

We note that dynamic critical exponents were recently reported by Canet and Moore \cite{KPZ-canet-moore} for two universality classes of the KPZ-equation.
For one type of approximate solution of the Flory-Imry-Ma or RSB-type, the dynamic exponent assumed the value $z=(4+d)/3$ below two dimensions, hence $z=5/3$ in $d=1$. Hence we state that the exact pseudo-dynamic critical exponent of RSB in the SK-model maps to the one of FIM or RSB-approximate solution of the KPZ-equation in 1D (provided one agrees to make the choice of $1/a$ being the pseudo-time which corresponds to the real time of KPZ). As in the DP-analogy, this should correspond to the domain-wall solution and hence to ${\cal CP}1$.

There is however also the known exact result of the 1D KPZ-equation $z=3/2$ also given by Canet and Moore\cite{KPZ-canet-moore}.

One may suspect that this result should be mappable to pseudo-dynamic behavior of RSB-SK near the second critical point ${\cal CP}2$.
Indeed, if we would conserve $z\hspace{.1cm}\nu=1$, the same exponent $z=3/2$ would be obtained near ${\cal CP}2$. We do not have any reason for this choice, and the explicit scaling of the order function near ${\cal CP}2$ does not confirm this value, neither for the choice $\tau=1/a$ nor for $\tau=a$. This question must remain open.
%XXXXXXXXXXXXXXXXXXXXXXXXXXXXXXXXXXXXXXXXXXXXXXXXXXXXXXXXXXXXXXXXXXXXXXXXXXXXXXXXXXXXXXXXXXXXXXXXXXXXXX
%XXXXXXXXXXXXXXXXXXXXXXXXXXXXXXXXXXXXXXXXXXXXXXXXXXXXXXXXXXXXXXXXXXXXXXXXXXXXXXXXX
\section{Detailed structure of the order function derivatives $q'(a)$ and $q''(a)$.}
\label{q(a)-derivatives}
\begin{figure}[here]
\resizebox{.48\textwidth}{!}{%
\hspace{-.2cm}
\includegraphics{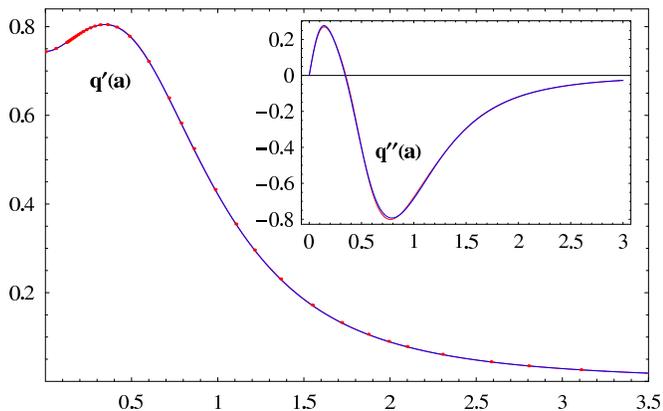}}
\caption{\label{fig:q-derivatives}The characteristic maximum in $q'(a)$ is analytically well reproduced. Red dots show the discretized derivative obtained from $50$ fixed point $\{a^*,q^*\}$. The insert shows the second derivative of the model order function $q(a)$.}
\end{figure}

%XXXXXXXXXXXXXXXXXXXXXXXXXXXXXXXXXXXXXXXXXXXXXXXXXXXXXXXXXXXXXXXXXXXXXXXXXXXXXXXXX
%
The derivatives depend much more specifically on the pseudo-time variable than $q(a)$ itself.
Failure of an analytic model function becomes detectable more easily in the derivatives. In order to control our modeling, we studied analytical fits first of all $200$-RSB data, and secondly of the $50$ calculated fixed points. Taking $q'(a)$ directly form the analytical form $q(a)$ as given by Eq.(\ref{model-function}), we find good agreement with the discretized slope calculated from the fixed points. This is demonstrated in the main part of Fig.\ref{fig:q-derivatives}.
In addition the insert shows the second derivative $\partial_a^2 q(a)$, where the two analytic models (red and blue curves) show a small difference. We note in passing that the shape of the 2nd derivative $q''(a)$ shows a similarity with the 2-loop correction in the $Y$-correlator of ($1+1$)-dimensional random bond pinned manifolds\cite{middleton} (we don't know whether this similarity has a deeper reason).
The maximum seen in $q'(a)$ expresses the Crisanti-Rizzo curvature\cite{Crisanti2002,prl2007}, a slight non-linearity of the order function in the small $a$ regime. It is however this contribution, which renders an analytical fit rather awkward. An analytical model which fits well the neighborhood of the critical points $a=0$ and $a=\infty$ can have a simpler shape\cite{prl2005}, but we want to get the pseudo-dynamic crossover right as well. Global quantities like the energy (integral over all $a$), picking up only small contributions nearby the critical points, depend on the crossover regime modeling. This can be seen in Eq.(\ref{gs-energy}) as well as in Fig.12 for the energy density.
%
%XXXXXXXXXXXXXXXXXXXXXXXXXXXXXXXXXXXXXXXXXXXXXXXXXXXXXXXXXXXXXXXXXXXXXXXXXXXXXX
\section{Conclusions}
%XXXXXXXXXXXXXXXXXXXXXXXXXXXXXXXXXXXXXXXXXXXXXXXXXXXXXXXXXXXXXXXXXXXXXXXXXXXXXX
%
In this article we formulated a scaling theory of the flow towards full replica symmetry breaking (RSB) at $T=0$, for finite temperatures, and for finite magnetic fields in the SK-model.
Several fixed point functions of RSB-flow were evaluated.

The analysis was guided by

1. a large set of high-precision numerical data, with up to $200$ self-consistently solved orders of replica symmetry breaking for the $T=0$ SK-spin glass and still a high number of orders for finite temperatures and magnetic fields,

2. by the identification of two critical points (at zero temperature and zero magnetic field), which are distinguished by two different pseudo-dynamic limits as obtained in an analytic picture of a Langevin-type equation in Ref.\onlinecite{prl2005,prl2007}, and

3. by representing nonanalytic behavior near each of these critical points in the framework of the scaling theory of critical phenomena.

Power laws and scaling functions were identified by fitting the leading $200$ RSB-orders of self-consistent solutions deep inside the SK spin glass phase; non-integer exponents were found and identified as rational numbers, characteristic of one-dimensional RSB-behavior. This 1D-character originates in correlations on the pseudo-lattice of RSB-orders $\kappa$. By means of scaling functions we demonstrated how these nonanalytic 1D-correlations enter in temperature- and field-dependent power laws in the ordered phase.

The universality class of replica symmetry breaking in the SK-model called for comparison with other physical systems, and shows similarities with directed polymers.

The decoupling of a magnetic field sensitive critical point from a temperature-sensitive one was embedded in an unconventional scaling hypothesis for the free energy and found to be consistent with the numerical data.

The RSB flow was used to generate an order parameter fixed point function, serving as a crossover between the two different pseudo-dynamical critical limits. Its fine structure was revealed by the leading derivatives, again confirming excellent agreement between analytical model and fixed point function.\\

\section{Acknowledgments}
We are indebted to Kay Wiese, Markus M\"uller, Thomas Garel, Andrea Crisanti, David Sherrington, Haye Hinrichsen, and Stefan Boettcher for stimulating discussions and helpful remarks.
We thank Tommaso Rizzo for useful remarks and for sending recent work prior to publication \cite{parisi-rizzo}.
We thank the DFG for partial and continued support of this research under grant Op28/7-1.

\end{document}